\documentclass[a4paper,11pt]{article}
\usepackage{jheppub} 
\usepackage{graphicx}
\usepackage{caption}
\usepackage{subcaption}
\usepackage{hyperref}
\usepackage[T1]{fontenc}
\usepackage{stfloats}
\usepackage{comment}
\usepackage{physics}
\usepackage{tensor}
\usepackage{placeins}
\usepackage{textcomp}
\usepackage{cuted}
\usepackage{xcolor}
\usepackage{amsfonts}
\usepackage{tikz}
\usetikzlibrary{matrix}

\usepackage{graphicx}%
\usepackage{multirow}%
\usepackage{amsmath,amssymb,amsfonts}
\usepackage{amsthm}%
\usepackage{mathrsfs}%
\usepackage[title]{appendix}%
\usepackage{xcolor}%
\usepackage{textcomp}%
\usepackage{manyfoot}%
\usepackage{booktabs}%
\usepackage{algorithm}%
\usepackage{algorithmicx}%
\usepackage{algpseudocode}%
\usepackage{listings}%

\def\CA{\mathcal{A}}
\def\CB{\mathcal{B}}

\def\CK{\mathcal{K}}

\title{\boldmath Pole-skipping of gravitational waves in the backgrounds of four-dimensional massive black holes}

\title{Pole-skipping of gravitational waves in the backgrounds of four-dimensional massive black holes}

\author[a,b]{Sa\v{s}o Grozdanov}
\author[a]{and Mile Vrbica}
\affiliation[a]{Higgs Centre for Theoretical Physics, University of Edinburgh, Edinburgh, EH8 9YL, Scotland}
\affiliation[b]{Faculty of Mathematics and Physics, University of Ljubljana, Jadranska ulica 19, SI-1000 Ljubljana, Slovenia}

\abstract{Pole-skipping is a property of gravitational waves dictated by their behaviour at horizons of black holes. It stems from the inability to unambiguously impose ingoing boundary conditions at the horizon at an infinite discrete set of Fourier modes. The phenomenon has been best understood, when such a description exists, in terms of dual holographic (AdS/CFT) correlation functions that take the value of `$0/0$' at these special points. In this work, we investigate details of pole-skipping purely from the point of view of classical gravity in 4$d$ massive black hole geometries with flat, spherical and hyperbolic horizons, and with an arbitrary cosmological constant. We show that pole-skipping points naturally fall into two categories: the algebraically special points and a set of pole-skipping points that is common to the even and odd channels of perturbations. Our analysis utilises and generalises (to arbitrary maximally symmetric horizon topology and cosmological constant) the `integrable' structure of the Darboux transformations, which relate the master field equations that describe the evolution of gravitational perturbations in the two channels. Finally, we provide new insights into a number of special cases: spherical black holes, asymptotically Anti-de Sitter black branes and pole-skipping at the cosmological horizon in de Sitter space.}

\begin{document}
\maketitle
\flushbottom

\section{Introduction}\label{sec:Intro}

Decades of studies of linearised perturbations around gravitational backgrounds have led to numerous physical insights into astrophysics and cosmology (see e.g.~Ref.~\cite{weinbergStevenWeinbergCosmology2009}), and with the help of the holographic duality (the AdS/CFT correspondence \cite{maldacenaLargeLimitSuperconformal1997}), more recently also into quantum field theories (QFTs) (see e.g.~Refs.~\cite{zaanenHolographicDualityCondensed2015,ammonGaugeGravityDuality2015,hartnollHolographicQuantumMatter2016}). With the theory of classical linearised gravitational perturbations so thoroughly investigated, it may seem unlikely that novel and potentially generic features of these equations could still be found. Nevertheless, recent holographic investigations of the connection between transport and quantum chaos in strongly coupled thermal quantum field theories uncovered such a phenomenon now known as {\em pole-skipping} \cite{grozdanovBlackHoleScrambling2018,blakeQuantumHydrodynamicalDescription2017,blakeManybodyChaosEnergy2018,grozdanovConnectionHydrodynamicsQuantum2018}. Even though pole-skipping originated from holography and is in a sense a statement about QFT correlation functions, the gravitational incarnation of it (actually, the better understood side of the phenomenon) requires no holography and is in itself worth studying. The new results on pole-skipping presented in this paper will focus almost exclusively on their role in classical gravity, within the realm of the theory of general relativity, and our analysis will rely solely on the vacuum Einstein equations.  

What is pole-skipping? Classical gravitational equations of motion for gauge-invariant (diffeomorphism-invariant) perturbations $\psi$ in Einstein gravity are second-order differential equations. This implies the existence of two independent solutions, which are, near the horizon, usually taken to be the ingoing and the outgoing modes. Since, classically, `everything' falls into a black hole, physical considerations normally require us to impose the ingoing boundary conditions at the black hole horizon. For certain parameters characterising the perturbations in Fourier space, however, there is no way to select a unique ingoing mode. In other words, two linearly independent ingoing modes exist. At each such special point --- and there is a discrete and infinite set of such points \cite{grozdanovComplexLifeHydrodynamic2019,blakeHorizonConstraintsHolographic2020} --- we therefore obtain an extra free parameter worth of `ingoing' solutions. These are the {\it pole-skipping} points, which signal an `indeterminacy'. The name is due to the fact that in the holographic description of this phenomenon, this indeterminacy at asymptotic infinity gives rise to a dual correlation function that (formally) has the value of `$0/0$' \cite{grozdanovBlackHoleScrambling2018,blakeQuantumHydrodynamicalDescription2017,blakeManybodyChaosEnergy2018}. Along a Fourier space `trajectory' of a specific pole of the correlator, there also lies a zero of the correlator that `cancels' the pole at the pole-skipping point. 

In the context of holography and maximally chaotic conformal field theories (CFTs) with a large number of local degrees of freedom (large-$N$ theories), i.e., in spaces with a negative cosmological constant, pole-skipping has been thoroughly investigated. Among various works that have analysed it are Refs.~\cite{grozdanovComplexLifeHydrodynamic2019,natsuumeHolographicChaosPoleskipping2019,haehlEffectiveFieldTheory2018, natsuumePoleskippingZeroTemperature2020,
wangPoleSkippingHolographic2022, natsuumeNonuniquenessGreenFunctions2019,natsuumeNonuniquenessScatteringAmplitudes2021, ahnClassifyingPoleskippingPoints2021, blakeHorizonConstraintsHolographic2020,blakeChaosPoleskippingRotating2022,liuQuantumChaosTopologically2020,Abbasi:2019rhy,Abbasi:2020ykq,Abbasi:2020xli,Amano:2022mlu,jansenQuasinormalModesCharged2020,ahnPoleskippingScalarVector2020,ahnScramblingHyperbolicBlack2019,grozdanovBoundsTransportUnivalence2020,Yuan:2023tft}. From the point of view of effective field theory, the symmetry principles that lead to pole-skipping were understood in Refs.~\cite{blakeQuantumHydrodynamicalDescription2017,Blake:2021wqj}. In general, what has been uncovered is that pole-skipping always occurs at frequencies $\omega$ that are integer multiples of the negative imaginary Matsubara frequency, i.e., $\omega = \omega_n$ \cite{grozdanovComplexLifeHydrodynamic2019,blakeHorizonConstraintsHolographic2020}, where the range of the integer $n$ depends on the spin of the operator \cite{wangPoleSkippingHolographic2022}. For bosonic operators, we have $\omega_n = -2 \pi i T n$ and for fermionic operators \cite{ceplakFermionicPoleskippingHolography2019,ceplakPoleskippingRaritaSchwingerFields2021}, we have $\omega_n = - 2 \pi i T (n+1/2)$. The wavevector associated with each of these (infinitely many) points can be in general complex. The leading pole-skipping point in the energy density correlator (the spin-$0$ component of the energy-momentum tensor), which was found as the first such case \cite{grozdanovBlackHoleScrambling2018,blakeQuantumHydrodynamicalDescription2017}, has been related to maximal quantum chaos \cite{shenkerBlackHolesButterfly2013,robertsLocalizedShocks2014,shenkerStringyEffectsScrambling2015,maldacenaBoundChaos2015} and can be used to compute the butterfly velocity of propagation of the chaotic wavefront with the Lyapunov exponent $\lambda_L = 2\pi T$. Beyond the semi-classical large-$N$ limit, pole-skipping is less well understood, but significant progress has also been made for such theories \cite{choiPoleSkippingAway2021}.\footnote{See also recent works on the effective field theory of non-maximal chaos \cite{Gao:2023wun,Choi:2023mab}.} 

In this work, we focus on studying and classifying pole-skipping points in four dimensional backgrounds of massive (non-rotating and uncharged) black holes with maximally symmetric 2$d$ horizons, and with an arbitrary cosmological constant. Our classification will be based on the special `integrable' structure of gravitational perturbation equations \cite{chandrasekharMathematicalTheoryBlack1983} that can be encoded in the language of Darboux transformations (for a recent work on this subject, see Ref.~\cite{lenziDarbouxCovarianceHidden2021}). While this structure has been known to exist for spherically symmetric asymptotically flat $4d$ black holes, to facilitate our classification, we generalise the formalism of Darboux transformations that relates the even and odd channels of linearised fluctuations to all considered black holes: asymptotically Minkowski, de Sitter and Anti-de Sitter black holes with flat, spherical and hyperbolic horizons. 

Should this be necessary, we stress that four-dimensional black holes are interesting for a number of reasons. Black holes in asymptotically Anti-de Sitter spaces are dual to 3$d$ CFTs in a thermal state. More importantly for this paper, the special integrable structure on which our classification is based is only known to exist in 4$d$. Furthermore, we appear to live in four (large) dimensions and there is little doubt left as to the existence of black holes in our universe. Finally, since observational data on gravitational waves is now readily available, a more thorough study of pole-skipping exhibited by 4$d$ gravitational waves may also play a role in future astrophysical and cosmological investigations. 

This paper is structured as follows: In Section~\ref{sec:BH}, we discuss the geometry of 4$d$ massive black holes and their gravitational perturbations in two channels distinguished by their parity transformations: the even and the odd channel. In Section~\ref{sec:Darboux}, we introduce the `integrable' structure encoded in the Darboux transformations that relate the gravitational wave solutions between the two channels. In Section~\ref{sec:PS}, we discuss general properties of pole-skipping and classify pole-skipping points of 4$d$ black holes. Then, in Section~\ref{sec:PS4}, we analyse a number of physically interesting special cases of our results. Moreover, since in asymptotically de Sitter spaces, cosmological horizons with analogous properties to those of the event horizons exist as well, we also examine properties of pole-skipping at the cosmological horizon. Finally, in Section~\ref{sec:Discussion}, we discuss a number of outstanding problems and outline some potential future research directions. 

\section{Massive $4d$ black holes and their perturbations}
\label{sec:BH}
\subsection{The background}
\label{subsec:BH_background}

In this work, we study static, 4$d$ solutions of the vacuum Einstein equations with an arbitrary cosmological constant $\Lambda$. In particular, we focus on the gravitational solutions, mainly black holes and black branes, expressed in the coordinate system $x^\mu = (t,r,\chi,\phi)$ for which the spatial constant-$r$ slices are maximally symmetric. Formally, we express all such geometries as the product space $\mathcal{M}\times\mathcal{K}$, where $\mathcal{M}$ is a $2d$ Lorentzian spacetime with the metric $g_{ab}$ and $\mathcal{K}$ a $2d$ maximally symmetric Riemannian space with the metric $\gamma_{AB}$:\footnote{
The Greek, the lower-case Latin and the upper-case Latin indices are used to run over the following coordinates: $\mu,\nu\in \{0,1,2,3\}$, $a,b \in \{0,1\}$ and $A,B \in\{2,3\}$, respectively.}
\begin{equation}
    \dd s^2 =g_{\mu\nu}\dd x^\mu \dd x^\nu = g_{ab}\dd x^a \dd x^b + r^2 \gamma_{AB} \dd x^A \dd x^B\label{metric},
\end{equation}
where
\begin{align}
    g_{ab}\dd x^a \dd x^b &= -f(r) \dd t^2 +\frac{1}{f(r)} \dd r^2, \\
    \gamma_{AB}\dd x^A \dd x^B &= \frac{\dd \chi^2}{1-K \chi^2}+\chi^2 \dd \phi.
\end{align}
The `emblackening factor' $f(r)$ is given by
\begin{equation}
    f(r)=K -\frac{2M}{r}-\frac{\Lambda}{3}r^2, \label{def:f}
\end{equation}
where $M$ is a constant representing the black hole's `mass', the Newton's constant is set to $G=1$ and $K$ is the normalised constant sectional curvature so that $K=1$, $K=0$ and $K=-1$ correspond to spherically, translationally and hyperbolically symmetric spaces, respectively. These three choices correspond to the well-known spherical, planar and hyperbolic black holes. As is usual, planar black holes will be referred to as black branes. The asymptotically flat spherical black hole is the simplest and the oldest known such solution dating back to 1916: the Schwarzschild black hole.

\begin{figure}[h!]
    \centering
    \includegraphics[scale=0.35]{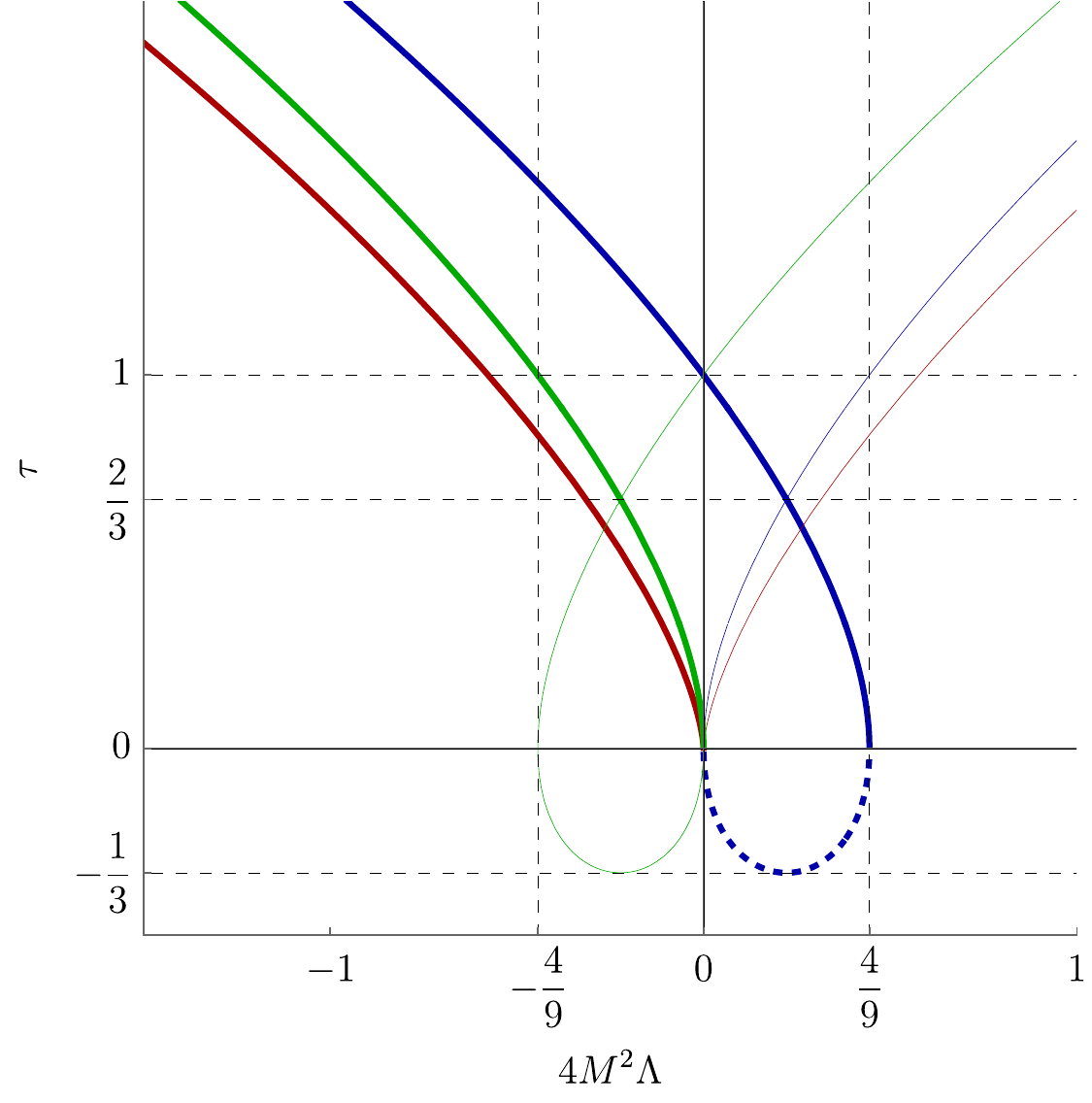}
    \caption{A representation of the solutions to Eq.~\eqref{taulambda} for spherical black holes with $K=1$ (blue), black branes with $K=0$ (red) and hyperbolic black holes with $K=-1$ (green). The event horizon branches are depicted with thick solid lines and the cosmological horizon with a dashed blue line. The event horizon branch is bijective for all physical $\tau$ while the cosmological horizon branch is not bijective. Black branes and hyperbolic black holes only exist in asymptotically AdS spaces with $\Lambda< 0$. Spherical black holes exist for any $\Lambda$.
    The unphysical solutions with $r_0<0$ (thin solid lines) will not be studied in this work.} 
    \label{fig:f}
\end{figure}

Black holes described by the spacetime metric \eqref{metric} have an event horizon at $r=r_0$, which is the smallest positive real root of the equation $f(r) = 0$. In the asymptotically flat ($\Lambda = 0$) and anti-de Sitter ($\Lambda < 0$) cases, this solution is the only physical solution of $f(r) = 0$. In the asymptotically de Sitter ($\Lambda > 0$) case, however, there may exist a second physical horizon, which is known as the cosmological horizon. An event horizon of a black hole has an associated Hawking temperature, which is given by
\begin{equation}\label{def:temperature}
T= \frac{f'(r_0)}{4\pi},
\end{equation}
where the prime ($'$) denotes a derivative with respect to the coordinate $r$. It will also prove convenient to define the following {\it normalised temperature}:  
\begin{equation}
\tau \equiv \frac{T}{T_0}=2Mf'(r_0)  \label{def:tau},
\end{equation}
where $T_0$ is the temperature of the asymptotically flat and spherically symmetric Schwarzschild black hole:
\begin{equation}
T_0 \equiv T(K=1,\Lambda=0)=\frac{1}{8\pi M}.
\end{equation}
We will parameterise the metric in terms of $\tau$ by using the identity
\begin{equation}
    \frac{\tau^2 (\tau-K^2)}{3}=16M^2 \Lambda \left( M^2 \Lambda-\frac{K}{9} \right), \label{taulambda}
\end{equation}
and express the position of the event horizon as
\begin{equation}
    r_0=2M \frac{\sqrt{K^2+3\tau}-K}{\tau}. \label{horizon}
\end{equation}
Note that in using the expression~\eqref{taulambda}, one must carefully choose the correct branches of the solutions. We plot the branches of physical solutions in the parameter space of $\tau$ and $4M^2\Lambda$ for $K = \{1,0,-1\}$ in Figure~\ref{fig:f}. If the 2$d$ manifold $\CK$ is either flat or negatively curved, i.e., for cases with $K=0$ (black brane) or $K=-1$ (hyperbolic black hole), the event horizon only exists when the 4$d$ geometry has $\Lambda < 0$ and is therefore asymptotically Anti-de Sitter. All such spaces with an event horizon have $\tau > 0$ (positive temperature). On the other hand, for positively curved $\CK$ with $K = +1$ (spherical black hole), physical solutions with (real) horizons also exist for non-negative values of the cosmological constant $\Lambda$. In terms of $\tau$, $\tau=1$ corresponds to all asymptotically flat black holes, $\tau>1$ to the asymptotically Anti-de Sitter (AdS) cases and $0<\tau<1$ to the asymptotically de Sitter (dS) cases. The space described by $\tau=0$ ($M^2 \Lambda = 1/9$) is called the Nariai space \cite{nariaiStaticSolutionsEinstein1950,nariaiNewCosmologicalSolution1951} and is the extremal solution with the overlapping event and cosmological horizons. For the parameter regime of $M^2\Lambda > 1/9$, the geometry has a naked singularity. Neither the Nariai space nor the cases with naked singularities will be of interest to us in this paper. We will therefore limit ourselves to geometries with $\tau > 0$. It should be noted that up to rescalings of the radial coordinate $r$, the choice of $K$ and $\tau$ completely parameterises all of the spaces considered here.

\subsection{Master perturbation equations}
\label{subsec:BH_mastereq}

Pole-skipping is a property of linearised gravitational perturbations in spacetimes with horizons. To facilitate the discussion of pole-skipping in the next section, we first present a general theory of perturbations of spacetimes described in Section~\ref{subsec:BH_background} in the language of the gauge-invariant {\it master fields} and {\it master equations} (see Refs.~\cite{kodamaMasterEquationGravitational2003,martelGravitationalPerturbationsSchwarzschild2005}). 

To write down the linearised perturbation equations of the background metric~\eqref{metric}, we start by introducing some notation. In particular, let $\nabla_a$ be the (unperturbed) covariant derivative on $\mathcal{M}$ defined with respect to the metric $g_{ab}$ and $D_A$ the (unperturbed) covariant derivative on $\mathcal{K}$ defined with respect to the metric $\gamma_{AB}$. We can then introduce a d'Alembertian on $\mathcal{M}$:
\begin{equation}
    \Box \equiv g^{ab} \nabla_a \nabla_b,
\end{equation}
and a Laplacian on $\mathcal{K}$:
\begin{equation}
    \Delta \equiv \gamma^{AB}D_A D_B.
\end{equation}
On $\mathcal{M}$, we also introduce a timelike Killing vector $t^a$ and a normalised radial vector $r_a$:
\begin{align}
    r_a &\equiv \frac{\partial r}{\partial x^a} , \\ t^a &\equiv -\epsilon^{ab}r_b,
\end{align}
where $\epsilon^{ab}$ is the 2$d$ Levi-Civita tensor. In terms of their components, we can set $r_t=0$, $r_r=1$, $t^t=1$ and $t^r=0$.

We can now perturb the metric to first order, $g_{\mu\nu}\rightarrow g_{\mu\nu}+\delta g_{\mu\nu}$, and decompose the fluctuations $\delta g_{\mu\nu}$ with respect to the symmetries (isometries) of the background. It is convenient to expand the metric perturbations in terms of the scalar eigenfunctions $Y^{(\mu,j)}$ of the Laplacian operator on the maximally symmetric 2$d$ spatial manifold $\mathcal{K}$:
\begin{equation}
   \left( \Delta +\mu \right) Y^{(\mu,j)}=0, \label{def:mu}
\end{equation}
were $j$ is an index labeling different eigenfunctions with the same eigenvalue. We will omit the index $j$ below. In this basis, the equations decouple. For more details, see Ref.~\cite{helgason2000groups}.

In the spherical case ($K=1$), these eigenfunctions are the spherical harmonics with a discrete spectrum of non-negative eigenvalues:
\begin{equation}\label{mu-l-forK1}
    \mu=\ell(\ell+1), ~~ \ell \in \mathbb{N}_0.
\end{equation}
In the planar case ($K=0$), the eigenfunctions are the Fourier modes with a continuous spectrum of non-negative eigenvalues:
\begin{equation}\label{mu-k-forK0}
    \mu=k^2,~~ k\in \mathbb{R},
\end{equation}
where $k$ is the wavevector. The formalism considered here and our results will hold also for the hyperbolic case ($K=-1$). For details on those eigenfunctions, we refer the reader to Ref.~\cite{helgason2000groups}. 

We can now use the scalar functions $Y^{(\mu)}$ on $\mathcal{K}$ to also introduce vectors and tensors on $\mathcal{K}$, which are needed to expand the metric (cf.~Ref.~\cite{sandbergblTensorSphericalHarmonics1978}):
\begin{subequations}
\begin{align}
    Y_A^{(\mu)}&=D_A Y^{(\mu)}, \\ X_A^{(\mu)}&=\epsilon{_A}{^B}Y_B^{(\mu)},  \\
    Y_{AB}^{(\mu)}&=D_A Y^{(\mu)}_B+\frac{1}{2}\mu Y^{(\mu)} \gamma_{AB},  \\ X_{AB}^{(\mu)}&=\frac{1}{2}\qty(D_A X^{(\mu)}_B+D_B X^{(\mu)}_A). 
\end{align}
\end{subequations}
Using these scalars, vectors and tensors, we can then decompose the perturbed metric as

\begin{subequations}
\label{g_perturbations}
\begin{align}
    \delta g^{(\mu)}_{ab}(t,r,\chi,\phi) &= h^{(\mu)}_{ab}(t,r) Y^{(\mu)}(\chi,\phi), \\
    \delta g^{(\mu)}_{aB}(t,r,\chi,\phi) &= j^{(\mu)}_a(t,r)Y^{(\mu)}_B(\chi,\phi)+v^{(\mu)}_a X^{(\mu)}_B(\chi,\phi), \\
    \delta g^{(\mu)}_{AB}(t,r,\chi,\phi) &= r^2 \qty[w^{(\mu)}(t,r) \gamma_{AB} Y^{(\mu)}(\chi,\phi)+q^{(\mu)}(t,r) Y^{(\mu)}_{AB}(\chi,\phi)]+s^{(\mu)}(t,r)X^{(\mu)}_{AB}(\chi,\phi),
\end{align}
\end{subequations}
where $h^{(\mu)}_{ab}$, $j^{(\mu)}_a$, $v^{(\mu)}_a$, $w^{(\mu)}$, $q^{(\mu)}$ and $s^{(\mu)}$, defined on $\mathcal{M}$, account for the ten independent components of the perturbed metric. We will henceforth also suppress the superscript $\mu$ as well as the explicit coordinate dependence. 

The equations of motion split into two independent channels: the even ($+$) channel that includes $h_{ab}$, $j_a$, $w$ and $q$, and the odd ($-$) channel that includes $v_a$ and $s$ (distinguished by how they transform under parity). By using
\begin{equation}
    \epsilon_a = j_a- \frac{1}{2}r^2\nabla_a q ,
\end{equation}
we can construct two gauge-invariant functions in the even channel:
\begin{subequations}\label{w_def}
\begin{align}
    \bar{h}_{ab}&=h_{ab}-\nabla_a \epsilon_b-\nabla_b \epsilon_a, \\
    \bar{w}&=w+\frac{\mu}{2}q-\frac{2}{r}r^a \epsilon_a,
\end{align}
\end{subequations}
and one gauge-invariant function in the odd channel:
\begin{equation}\label{v_def}
    \bar{v}_a=v_a-\frac{1}{2}\nabla_a s+\frac{r_a}{r}s.
\end{equation}
Furthermore, for future use, we also define a function of the radial coordinate $\Phi(r)$:
\begin{equation}
    \Phi(r)=\mu+r f'(r)-2f(r)=\mu-2K+\frac{6M}{r}. \label{Phi}
\end{equation}
Although not apparent at the moment, the behaviour of this function will be of crucial importance for the results of this paper.

Using these gauge-invariant variables and their first derivatives, one can now define the gauge-invariant {\it master functions} in the even ($\psi_+$) and the odd ($\psi_-$) channel as follows \cite{zerilliEffectivePotentialEvenparity1970, zerilliGravitationalFieldParticle1970, moncriefGravitationalPerturbationsSpherically1974,chandrasekharEquationsGoverningPerturbations1975, sQuasinormalModesSchwarzschild1975,reggeStabilitySchwarzschildSingularity1957,cunninghamRadiationCollapsingRelativistic1978, cunninghamRadiationCollapsingRelativistic1979}:
\begin{subequations}
\label{def:master}
\begin{align} \psi_+&=r \qty[\bar{w}+\frac{2}{\Phi(r)}\qty(r^a r^b \bar{h}_{ab}-r r^a \nabla_a \bar{w})], \\
    \psi_-&=r \epsilon^{ab}\qty(\nabla_a \bar{v}_b-\frac{2}{r}r_a\bar{v}_b).
\end{align}
\end{subequations}  
If we know the master function, then, by using the perturbed Einstein equations, we can also recover all components of the perturbed metric $\delta g_{\mu\nu}$ (see e.g.~Ref.~\cite{kodamaMasterEquationGravitational2003}). For a comprehensive review of the master function formalism in 4$d$, see Ref.~\cite{lenziMasterFunctionsEquations2021}.

The gauge-invariant master functions $\psi_\pm$ satisfy the so-called {\it master equations}: 
\begin{equation}
    \left( \Box-V_\pm(r) \right) \psi_\pm(t,r)=0 . \label{mastereq}
\end{equation}
The two respective potentials $V_\pm(r)$ are given by
\begin{subequations}
\label{potentials}
\begin{align}
    V_+(r)&=\frac{72 M^3-12 M^2 r \qty[ 2 \Lambda  r^2 -3 (\mu-2K)]+6 M r^2 (\mu -2 K )^2+  r^3 \mu(\mu -2 K )^2}{r^3 \qty[6 M+r (\mu -2 K )]^2}, \label{Veven} \\
    V_-(r)&=\frac{\mu  r-6 M}{r^3}. \label{Vodd} 
\end{align}
\end{subequations}
In Section~\ref{sec:Darboux}, we will elucidate the rich hidden symmetry structure that relates the two potentials, encoded in the Darboux transformations, which is analogous to certain integrable structures known in the theory of differential equations.

Eqs.~\eqref{mastereq} can be further reduced by using the time translation symmetry of the backgrounds. We Fourier decompose the time dependence as
\begin{equation}
\psi_\pm(t,r) =  \psi_\pm(r)e^{-i\omega t} \label{fourier},
\end{equation}
where $\omega$ is the frequency. The master equations \eqref{mastereq} can then be written as 
\begin{equation}
    \qty(\frac{\dd^2}{\dd r_*^2}+\omega^2)\psi_\pm(r)=f(r)V_\pm (r)\psi_\pm(r). \label{mastereq2}
\end{equation}
Hence, we have reduced the entire problem of first-order gravitational perturbations of static black hole backgrounds to a one-dimensional scattering problem written as a Schr\"{o}dinger equation. Here, have also introduced the `tortoise coordinate' as
\begin{equation}
    \frac{\dd r_*}{\dd r}=\frac{1}{f(r)}. \label{def:tortoise}
\end{equation}
Asymptotically, near the boundary of AdS, near infinity of the Minkowksi space or near the cosmological horizon of dS, the tortoise coordinate behaves as  $r_* \to \infty$. Near the black hole event horizon $r = r_0$, $r_*$ scales as  
\begin{equation}
    r_* \sim \frac{1}{4\pi T} \ln(r-r_0), \label{tortoise_log}
\end{equation} 
which implies that $r_* \to -\infty$. Since the product $f(r) V(r)$ falls exponentially as $r_* \to -\infty$, it follows that near the horizon, the master equations simplify to 
\begin{equation}\label{master_horizon}
    \qty(\frac{\dd^2}{\dd r_*^2}+\omega^2)\psi_\pm(r)=0.
\end{equation}
It is important to note that the above master equation formalism breaks down when $\mu(\mu-2K)=0$ and when $\omega = 0$. Each of those cases will be studied independently. We also note that for $\mu=0$, the first-order perturbation itself exhibits the same symmetry as the spacetime background and is a static perturbation of the background geometry. It can be understood as a `perturbation' of the constant `mass' parameter $M$. The second case, $\mu=2K$, only exists for spherical horizons with $K=1$ and is, given Eq.~\eqref{mu-l-forK1}, therefore the spherical harmonics mode with $\ell = 1$. Thus, again, the perturbation is static. In the odd channel, it is the perturbation of the angular momentum and in the even channel, it can be understood as a boost of the coordinate system. For further details, see Refs.~\cite{martelGravitationalPerturbationsSchwarzschild2005,nakamuraGaugeinvariantPerturbationTheory2021,nakamuraGaugeinvariantPerturbationTheory2021a, nakamuraGaugeinvariantPerturbationTheory2021b}. All perturbations with $\ell \geq 2$ can be treated directly in the master field formalism. Finally, note that we will treat $\mu$ as a complex variable throughout most of the discussion.

\subsection{Asymptotics and boundary conditions}
\label{subsec:BH_asymptotics_BCs}

The master equations \eqref{mastereq2} are ordinary second-order differential equations. Each equation therefore has two independent solutions and requires a choice of two boundary conditions. In general, we can express these two solutions in an independent manner in a way that separates their behaviour asymptotically far from the horizon (at infinity or at the cosmological horizon, depending on $\Lambda$) and at the horizon. Asymptotically far, we write (each) $\psi$ as
\begin{equation}
    \psi = A\psi^\text{A}+B\psi^\text{B}, \label{def:AB}
\end{equation} 
where $A$ and $B$ are arbitrary constants. At the horizon $r=r_0$, we can instead write
\begin{equation}\label{PsiHorizon}
    \psi= C \psi^\text{in}+D\psi^\text{out},
\end{equation}
where, now, $C$ and $D$ are arbitrary constants (that can be related to $A$ and $B$, and vice versa). Near the horizon, the two independent solutions are the ingoing and the outgoing solution. 

By using Eq.~\eqref{master_horizon}, we see that, at the horizon, the ingoing and the outgoing solutions behave as
\begin{align}
    \psi^\text{in}(r \rightarrow r_0)&\sim e^{-i \omega r_*}\sim (r-r_0)^{-\frac{i\omega}{4\pi T}}, \\ 
    \psi^\text{out}(r \rightarrow r_0)&\sim e^{i \omega r_*}\sim (r-r_0)^{\frac{i\omega}{4\pi T}}. \label{around_horizon}
\end{align}
Note that by taking $\omega\rightarrow -\omega$, we map an ingoing solution to an outgoing one, and vice versa. The result in Eq.~\eqref{around_horizon} is `universal' in the sense that it is dictated by the horizon geometry and independent of any other details characterising the background, such as $K$ or $\Lambda$. 

Asymptotically far from the horizon, the scaling of solutions depends on the curvature of the spacetime. In the asymptotically flat case ($\Lambda = 0$), $r_* \sim r$, so the product $f(r)V(r) \to 0$ as $r\to\infty$ and the two solutions behave as
\begin{subequations}
\label{around_infinity}
\begin{align}
    \psi^A_\text{flat}(r \rightarrow \infty)&\sim e^{-i \omega r}, \\
    \psi^B_\text{flat}(r \rightarrow \infty)&\sim e^{i \omega r}. \label{around_infinity_flat}
\end{align}
When the space is asymptotically AdS ($\Lambda < 0$), we have 
\begin{align}
    \psi^A_\text{AdS}(r \rightarrow \infty)&\sim 1+\mathcal{O}(r^{-2}), \\ \psi^B_\text{AdS}(r \rightarrow \infty)&\sim \frac{1}{r}+\ldots. \label{AdS_asymptotics}
\end{align}
In the asymptotically dS case ($\Lambda > 0$), the asymptotic geometry has a  cosmological horizon $r_c$ (where, also, $r_*\rightarrow \infty$). In analogy with the behaviour of $\psi$ at the (standard black hole) event horizon, the solutions are also the ingoing and outgoing modes:
\begin{align}
    \psi^A_\text{dS}(r\to r_c)&\sim e^{-i \omega r_*}\sim (r_c-r)^{-\frac{i\omega}{f'(r_c)}}, \\ \psi^B_\text{dS}(r\to r_c)&\sim e^{i \omega r_*}\sim (r_c-r)^{\frac{i\omega}{f'(r_c)}}.
\end{align}
\end{subequations}
As at an event horizon, we can again map an ingoing solution to an outgoing one, and vice versa, by taking $\omega \to - \omega$.

With the knowledge of the asymptotic behaviour of $\psi$ in hand, we can now discuss the appropriate boundary conditions. Since the event horizon of a black hole classically absorbs all information, it is standard to impose ingoing boundary conditions at the horizon: i.e., $D = 0$ (cf.~Eq.~\eqref{PsiHorizon}). Such a boundary condition fixes the ratio of $B/A$ at asymptotic infinity, as it does the ratios of $C/A$ and $C/B$, which are important for the time evolution of gravitational waves.. In the language of the holography, this choice gives rise to the {\it retarded} dual two-point function of the energy-momentum tensor that is determined by $B/A$ \cite{sonMinkowskispaceCorrelatorsAdS2002,Herzog:2002pc}. Crucially, however, there exist special cases --- the pole-skipping points --- where one cannot uniquely impose the ingoing boundary conditions and end up with a single mode $\psi^\text{in}$ by setting $D=0$  \cite{grozdanovBlackHoleScrambling2018,blakeManybodyChaosEnergy2018}. 

At infinity, for asymptotically Minkowski and dS black holes, a standard choice is to set $B = 0$, which only keeps the outgoing modes --- no perturbation comes from infinity. The resulting solutions are called the {\it quasinormal modes} (QNMs), which are central to the analysis of gravitational waves. One could alternatively keep only the ingoing modes at infinity by using what is called the `totally transmissive' condition and set $A=0$. In asymptotically AdS spaces, which are compact, one usually either imposes the Dirichlet or the Neumann boundary conditions. What we typically refer to as the QNMs in asymptotically AdS spaces are the solutions with the Dirichlet boundary conditions and $A = 0$. In the language of holography, if the dual correlator is determined by the ratio of $B/A$, the QNM solutions with $A=0$ directly determine the poles of the corresponding correlator \cite{kovtunQuasinormalModesHolography2005}. Note, however, that when using the master fields, the relation between these conditions and the QNMs for pure metric perturbations $\delta g_{\mu\nu}$ is more involved. For a further discussion of QNMs, see e.g.~the review \cite{bertiQuasinormalModesBlack2009}.

\section{Darboux transformations}
\label{sec:Darboux}
\subsection{The formalism}

A remarkable fact about the linearised fluctuations of black holes in 4$d$ Einstein gravity, which was described by Chandrasekhar for asymptotically flat spaces in Refs.~\cite{chandrasekharEquationsGoverningGravitational1976,chandrasekharAlgebraicallySpecialPerturbations1984, chandrasekharMathematicalTheoryBlack1983}, is a special symmetry transformation that relates the even and odd channels. There exists a relation between the two potentials in Eq.~\eqref{potentials}, which implies that given a solution of the master equation in one of the channels, we can directly find a solution in the other channel (for the same values of $\omega$ and $\mu$) by using a first-order differential operator. This transformation between the channels, sometimes referred to as the Chandrasekhar transformation, is a special case of the Darboux transformations, introduced already in the late nineteenth century \cite{darbouxPropositionRelativeLinear1999}. While widely studied in the context of exactly solvable models in quantum mechanics, integrability and supersymmetry (see e.g.~Refs.~\cite{cooperSupersymmetryQuantumMechanics1995, matveevDarbouxTransformationsSolitons1991, guDarbouxTransformationsIntegrable2005, liuDarbouxTransformationsSUSY1998, matveevDarbouxTransformationsIntegrable, purseyIsometricOperatorsIsospectral1986}), its application to the theory of black holes was noticed only later \cite{glampedakisDarbouxTransformationBlack2017}.

The formalism of Darboux transformations allows us to introduce the notion of the so-called {\it algebraically special} modes and frequencies (see Refs.~\cite{anderssonTotalTransmissionSchwarzschild1994,brinkAnalyticTreatmentBlackhole2000, cardosoStabilityNakedSingularities2006, glampedakisDarbouxTransformationBlack2017, lenziDarbouxCovarianceHidden2021, yurovLookGeneralizedDarboux2019,leungSUSYTransformationsQuasinormal1999,lenziBlackHoleGreybody2022, rosenBlackHolePerturbations2021,headingResolutionMysteryChandrasekhar1977,Bakas_2009,Miranda_2006}). Even though we only consider massive black holes here, the formalism can also extended to the study of charged \cite{andersonIntertwiningEquationsBlackhole1991} and rotating \cite{chandrasekharEquationsGoverningGravitational1976,chandrasekharAlgebraicallySpecialPerturbations1984, glampedakisDarbouxTransformationBlack2017} black holes. It is believed that the rich structure embodied by the formalism of Darboux transformations is unique to black holes in 4$d$ \cite{ishibashiPerturbationsStabilityStatic2011}. 

In this work, we present this formalism in what we believe is its most elegant incarnation, generalising it to cases with an arbitrary cosmological constant $\Lambda$ and maximally symmetric horizon topology $K$ --- i.e., for all backgrounds that were introduced in Section~\ref{sec:BH}. We begin by writing the potentials $V_\pm(r)$ in the following form:
\begin{equation}
    f(r)V_\pm(r)=\tilde\omega^2+W(r)^2\pm\frac{\dd W(r)}{\dd r_*}, \label{potentialsDarboux}
\end{equation}
where
\begin{equation}
    W(r)=-\frac{\dd \ln{\Phi(r)}}{\dd r_*} -i\tilde\omega. \label{def:W}
\end{equation}
In this expression, $\Phi(r)$ is the function defined in Eq.~\eqref{Phi} and we introduced the frequency $\tilde\omega$:
\begin{equation}
   \tilde\omega=i\frac{\mu\left(\mu-2K\right)}{12M}. \label{omegastar}
\end{equation}
Next, we define two linear first-order differential operators $L_\pm$ (sometimes referred to as the {\it supersymmetry operators}):
\begin{equation}
    L_\pm=W(r)\pm\frac{\dd}{\dd r_*}. \label{def:L}
\end{equation}
The master equations \eqref{mastereq2} in the two channels can then be written as
\begin{subequations}
\label{mastereq3}
\begin{align}
    L_+ L_- \psi_+&=\left(\omega^2-\tilde\omega^2\right) \psi_+, \\
    L_- L_+ \psi_-&=\left(\omega^2-\tilde\omega^2\right) \psi_-.
\end{align}
\end{subequations}

Using the equations in the form expressed in Eq.~\eqref{mastereq3}, it is now follows that given a solution $\psi_+$ in the even channel, a new function 
\begin{equation}
\psi_-=L_- \psi_+    
\end{equation} 
will automatically solve the odd channel equation. Similarly, given a solution $\psi_-$ of the odd channel equation, the function 
\begin{equation}
\psi_+=L_+ \psi_-
\end{equation} 
will solve the even channel fluctuation equation. Furthermore, the transformation maps ingoing (or outgoing) waves at the horizon from one channel to ingoing (or outgoing) waves at the horizon in the other channel. In the asymptotically flat case, the same can be said about the ingoing and outgoing nature of the waves at asymptotic infinity, which is the property that yields the isospectrality (a one-to-one map) of QNMs in the two channels. As can be inferred from our above generalisation to spacetimes with different asymptotics, isospectrality is no longer guaranteed when $\Lambda \neq 0$.\footnote{Note that isospectrality can still hold in asymptotically dS spaces, but that depends on how precisely we define QNMs.}

\subsection{Algebraically special solutions}
\label{subsec:Darboux_AlgSpec}

The above discussion has important exceptions. In particular, since the operators $L_\pm$ are first-order operators, they can have a non-trivial kernel. This means that there exist functions that are directly annihilated by these operators. We will denote such special functions by $\tilde\psi_\pm (r)$. More precisely, $\tilde\psi_\pm (r)$ satisfy the following equations:
\begin{align}
    L_- \tilde\psi_+&=0, \\ L_+ \tilde\psi_-&=0. \label{alg_spec}
\end{align}
While $\psi_\pm^*$ defined in this way do not solve the master equations \eqref{mastereq3} in general, they do solve them for special values of frequencies: $\omega^2=\tilde\omega^2$. We will call such frequencies the {\it algebraically special frequencies} and the solutions \eqref{alg_spec} the {\it algebraically special solutions} (or modes). Moreover, we will call the pairs $(\omega,\mu)$ with $\omega^2=\tilde\omega^2$ the {\it algebraically special points}. Explicitly, the algebraically special solutions are given by
\begin{align}
    \tilde\psi_+&=\qty[\Phi(r) \, e^{ i\tilde\omega r_*}]^{-1} ,\\
    \tilde\psi_-&=\qty[\Phi(r) \, e^{ i\tilde\omega r_*}]^{+1} .\label{def:AS}
\end{align}

Setting $\omega^2=\tilde\omega^2$ does not automatically imply that a corresponding solution to the master equation will be an algebraically special solution. This is because there always exists a second set of linearly independent solutions (one even and one odd), call them $\tilde\chi_\pm(r)$, which are not annihilated by $L_\pm$. Since it can be shown that 
\begin{equation}
    L_- \tilde\chi_+=\tilde\psi_- \qquad L_+ \tilde\chi_-=\tilde\psi_+,
\end{equation}
we can also find explicit solutions for $\chi_+^*$ and $\chi_-^*$:
\begin{subequations}
\label{chis}
\begin{align}
    \tilde\chi_+&=\tilde\psi_+ \int\dd r_* \frac{\tilde\psi_-}{\tilde\psi_+}=\qty[\Phi(r)e^{ i\tilde\omega r_*}]^{- 1} \int \dd r_* \qty[\Phi(r)e^{ i\tilde\omega r_*}]^{+2} , \\ 
    \tilde\chi_-&=\tilde\psi_- \int\dd r_* \frac{\tilde\psi_+}{\tilde\psi_-} =\qty[\Phi(r)e^{ i\tilde\omega r_*}]^{+1} \int\dd r_* \qty[\Phi(r)e^{ i\tilde\omega r_*}]^{-2}, 
\end{align}
\end{subequations}
where the integration constants are chosen so that we get an ingoing (outgoing) $\tilde\chi$ for an outgoing (ingoing) $\tilde\psi$. 

The form of the above equations is indicative of the fact that an important subtlety related to the classification of ingoing and outgoing modes arises if the function $\Phi(r)$ (cf.~Eq.~\eqref{Phi}) vanishes. And as it turns out, $\Phi(r)$ can indeed vanish at the horizon. In terms of $\mu$, the equation $\Phi(r_0)=0$ is solved when
\begin{equation}
   \mu=\bar\mu \equiv 2K-\frac{6M}{r_0}=-4\pi T r_0=K-\sqrt{K^2+3\tau}. \label{mu0}
\end{equation}
The corresponding value of $\tilde\omega$ that follows from Eq.~\eqref{def:W} is then $\tilde\omega = 2\pi T i$. We will discuss the physical significance of this value in Section~\ref{subsec:PS4_AlgSpec}. We know that for `generic' $\mu$ (giving some $\omega=\tilde\omega$), the even master field solution is ingoing and the odd solution is outgoing at the horizon. If $\omega=-\tilde\omega$, the opposite is true. If, however, $\mu = \bar\mu$ so that $\Phi(r_0)=0$, then $e^{i\tilde\omega r_*}\sim (r-r_0)^{-1/2}$ and since $\Phi(r)\sim \Phi'(r_0)(r-r_0)$, we have
\begin{equation}
    \Phi(r)e^{i\tilde\omega r_*} \sim \Phi'(r_0)(r-r_0)^{1/2}.
\end{equation}
Hence, the nature of the ingoing versus the outgoing modes is reversed as compared to those which have $\Phi(r_0) \neq 0$. We summarise these properties in Table~\ref{table}, where it should be emphasised that the details of this discussion are specific to the master function formalism and do not directly apply to the ingoing/outgoing properties of the metric perturbations themselves.\footnote{We note that for solutions with $\Phi(r_0) \neq 0$ in asymptotically flat spacetimes ($\Lambda = 0$), the ingoing algebraically special solutions are ingoing also at asymptotic infinity. This means that for `generic' values of $\omega$, such examples are totally transmissive \cite{anderssonTotalTransmissionSchwarzschild1994}. This is, however, no longer true for the algebraically special solutions with $\mu = \bar \mu$, which have $\Phi(r_0) = 0$.} 

\begin{table}[ht!]
\centering
\begin{tabular}{|c| c c|c c |}
    \hline
    & \multicolumn{2}{|c|}{$\Phi(r_0)\neq 0$} &\multicolumn{2}{|c|}{$\Phi(r_0)=0$}\\
    \hline & $\omega=\tilde\omega$&$\omega=-\tilde\omega$  & $\omega=\tilde\omega$& $\omega=-\tilde\omega$ \\ \hline
   even ($+$)  & ingoing & outgoing  & outgoing & ingoing\\
   odd ($-$) & outgoing & ingoing &ingoing & outgoing    \\\hline
\end{tabular}
\caption{Summary of the behaviour of the algebraically special solutions at the event horizon for $\Phi(r_0) \neq 0$ ($\mu \neq \bar\mu$) and $\Phi(r_0) = 0$ ($\mu = \bar\mu$).
}
\label{table}
\end{table}

\section{Pole-skipping in $4d$ black hole backgrounds}
\label{sec:PS}
\subsection{Definition, classification and elementary properties}
\label{subsec:PS_def_class}

One way to think of pole-skipping, which was originally discovered in the context of holography and holography-inspired effective field theories~\cite{grozdanovBlackHoleScrambling2018,blakeQuantumHydrodynamicalDescription2017,blakeManybodyChaosEnergy2018,grozdanovConnectionHydrodynamicsQuantum2018}, is as a property of quantum field theory correlation functions. While the manifestation of the phenomenon is indeed perhaps clearest in QFT: pole-skipping points are points in the frequency-wavevector $(\omega,k)$ space where the correlator takes the value of `$0/0$', its interpretation is best understood from the dual (holographic), gravitational perspective. Moreover, in itself, the phenomenon requires no reference to a dual QFT. It can be thought of as purely a property of (linearised) gravitational waves in black hole spacetimes with event horizons. This is the perspective that we take in this paper. 

We start by defining what precisely we mean by pole-skipping in the context of gravity. Mostly, we will work directly with the gauge-invariant master equations. Recall that in Section~\ref{subsec:BH_asymptotics_BCs}, we discussed how the choice of (ingoing) boundary conditions at the horizon {\it uniquely} fixes the ratio of $B/A$, where $A$ and $B$ parametrise the two independent solutions at the boundary (cf.~Eq.~\eqref{def:AB}). However, it turns out this is not always possible. There exists an infinite sequence of points  $(\omega,\mu)$ for which imposing the ingoing boundary condition at the horizon \emph{does not} determine the solution uniquely. In other words, despite such boundary conditions, one still retains two linearly independent ingoing solutions. This results in a free parameter and makes the ratio of $B/A$ formally `$0/0$' in the sense that, while usually finite, $B/A$ is now a multivalued function at those points and its value depends on the limit of approach in the $(\omega,\mu)$ space. This is {\it pole-skipping}. Here, we will examine the gravitational details of this mechanism  and classify all such points $(\omega,\mu)$ for 4$d$ black holes introduced in Section~\ref{sec:BH} from the viewpoint of Darboux transformations discussed in Section~\ref{sec:Darboux}.

Consider the master equations~\eqref{mastereq2}. For both, even and odd channels, the equation is a second-order ordinary differential equation (ODE):
\begin{equation}
    f(r) \psi''(r)+f'(r)\psi'(r)+\qty(\frac{\omega^2}{f(r)}-V(r))\psi(r)=0. \label{master_analysed}
\end{equation}
In the language of the Frobenius-Fuchs theory, the event horizon and, when it exists, the cosmological horizon are regular singular points of the ODE, denoted by $r=r_0$. This means that a local expansion of $\psi(r)$ around $r_0$ requires us to factor out a singular piece, $\psi(r) = (r-r_0)^b g(r)$, where $b$ can take two values (the ODE is of 2nd order) and $g(r)$ is analytic at $r_0$. Hence, it can be expanded in a Taylor series. Depending on the exact details of the equation, when the two solutions for the exponent $b$ are separated by an integer, one of the two solutions for $g(r)$ can also be accompanied by a logarithm. Here, we will state the relevant results from the Frobenius-Fuchs theory for the case at hand. For details, see Refs.~\cite{coddington1955theory,bender1999advanced}.

Let $g_\text{in}(r)$ and $g_\text{out}(r)$ be the two analytic functions at $r_0$ and $g_\text{in}(r_0)\neq 0$ and let $g_\text{out}(r_0)\neq 0$. The analysis of Eq.~\eqref{master_analysed}, which gives $b = \pm i \omega / (4\pi T)$, where $T$ is the Hawking temperature of the horizon then results in the following possible cases that are distinguished by whether the frequencies are integer multiples of the negative imaginary Matsubara frequency, i.e., 
\begin{equation}\label{Matsubara}
    \omega_n = - 2 \pi T i n, ~~ n \in \mathbb{Z},
\end{equation}
or not. We note that all pole-skipping points have $n \geq -1$.

Given our above discussion of the exponents $b$, we consider the following two cases:
\begin{enumerate}
    \item ${i \omega}/{2\pi T} \notin \mathbb{Z}$
    \begin{align}
        \psi_\text{in}(r)&=(r-r_0)^{-\frac{i \omega}{4\pi T}}g_\text{in}(r), \\ \psi_\text{out}(r)&=(r-r_0)^\frac{i \omega}{4\pi T}g_\text{out}(r). \label{nonsingular}
    \end{align}
    The two solutions are the standard ingoing and outgoing waves and no pole-skipping occurs.
    
    \item ${i \omega}/{2\pi T} \in \mathbb{Z}$
    
    \begin{enumerate}
    \item Any case with $\omega=0$ (i.e., $n=0$) is a static perturbations, which cannot be accounted for by the formalism described in Section~\ref{subsec:BH_mastereq}. One must perform an independent analysis to see whether any pole-skipping points with $\omega = 0$ exist.\footnote{Note that, for example, in holographic duals of black branes, the $(k=0,\omega = 0)$ point is a `hydrodynamic' pole-skipping point. See Ref.~\cite{grozdanovComplexLifeHydrodynamic2019}.}
    \item If $i\omega/2 \pi T >0$, then the solutions are
    \label{ingoingPS}
    
    \begin{subequations}

    \label{apparent}
    \begin{align}
        \psi_\text{in}(r)&=(r-r_0)^{-\frac{i\omega}{4 \pi T}}g_\text{in}(r)+c \ln(r-r_0) \psi_\text{out}(r), \\ \psi_\text{out}(r)&=(r-r_0)^{\frac{i\omega}{4 \pi T}}g_\text{out}(r).  \label{apparent1}
    \end{align}

    Here $c=c(\omega,\mu)$ is a constant that can also be zero. If $c=0$, then we call $r=r_0$ an \emph{apparent singularity} of the differential equation. These points give rise to {\it almost all} pole-skipping solutions.
    
    \item \label{outgoingPS} If $i\omega/2 \pi T <0$, then the solutions are
    \begin{align}
        \psi_\text{in}(r)&=(r-r_0)^{-\frac{i\omega}{4 \pi T}}g_\text{in}(r),\\ \psi_\text{out}(r)&=(r-r_0)^{\frac{i\omega}{4 \pi T}}g_\text{out}(r)+c \ln(r-r_0) \psi_\text{in}(r).  \label{apparent2}
    \end{align}
        \end{subequations}
    This case becomes identical to \ref{ingoingPS} upon the following replacements: $\omega \rightarrow -\omega$ and $\psi_\text{in} \leftrightarrow \psi_\text{out}$.
    \end{enumerate}
\end{enumerate}

Let us now focus on the second family of cases with ${i \omega}/{2\pi T} \in \mathbb{Z}$, which can give rise to the sought pole-skipping modes. Since these cases mainly fall into the category~\ref{ingoingPS}, we start by using the solutions in Eq.~\eqref{apparent1}. Let us consider points $(\omega,\mu)$ that give $c(\omega,\mu)=0$. Having `cancelled' the logarithm, we have
\begin{align}
    \psi_\text{in}(r)&=(r-r_0)^{-\frac{n}{2}}g_\text{in}(r), \\ \psi_\text{out}(r)&=(r-r_0)^{\frac{n}{2}}g_\text{out}(r).
\end{align}
In this case, $\psi_\text{in}(r)$ is not the only function that satisfies the ingoing boundary condition. To see this, consider another solution $\psi^a_\text{in}(r)$ defined in the following way:
\begin{align}
    \psi^a_\text{in}(r)&=(r-r_0)^{-\frac{n}{2}}\qty[g_\text{in}(r)+a(r-r_0)^n g_\text{out}(r)] =(r-r_0)^{-\frac{n}{2}} \tilde{g}^a_\text{in}(r). \label{free_parameter}
\end{align}
The function $\tilde{g}^a_\text{in}(r)$ is analytic (which would not be the case if $c\neq 0$), with $\tilde{g}^a_\text{in}(r_0)\neq 0$, therefore $\psi^a_\text{in}(r)$ represents a legitimate ingoing solution for \emph{any} choice of $a$. In other words, fixing the ingoing boundary condition at the horizon fails to uniquely determine the solution of the master equation --- we are left with a free parameter. We will call such a choice of $(\omega,\mu)$ an {\it (ingoing) pole-skipping point}. Note that this argument also shows the necessity for `cancelling' the logarithms and therefore for having $c(\omega,\mu) = 0$.

Due to the fact that there exists a symmetry between the odd and the even channel, it turns out that both channels share an infinite set of {\it common} pole-skipping points. To see this, consider Darboux transforming a solution of Eq.~\eqref{free_parameter} in any one of the channels. If the original solution is {\it not} an algebraically special solution, then the Darboux transformed solution is a solution of the master equation in the other channel. It has the form
\begin{equation}
L_\pm\psi_\text{in}^a=(r-r_0)^{-\frac{n}{2}}\tilde{h}^a_\text{in}(r),
\end{equation} 
where $\tilde{h}^a_\text{in}(r)$ is an analytic function at $r_0$ with a free parameter $a$. Since the values of $\omega$ and $\mu$ are unaffected by this transformation and $\tilde{h}^a_\text{in}(r)$ still includes the free parameter $a$, this means that we have a pole-skipping point with the same $\omega$ and $\mu$ in the other channel as well. 

If, instead, a specific pole-skipping solution is an ingoing algebraically special pole-skipping solution, then we can write the solution (say, in the even channel) as
\begin{equation}
    \psi_+(r)=\tilde\psi_+(r)+a\tilde\chi_+(r), \label{AS_form}
\end{equation}
where $\tilde\psi_+$ is ingoing.
Using the defining property of the algebraically special modes, it follows that
\begin{equation}
    L_- \psi_+(r)=a L_-\tilde\chi_+(r)=a\tilde\psi_-(r),
\end{equation}
which does not have the pole-skipping form. Moreover, given the same $\omega$, $\tilde\psi_-$ will be ingoing. The same statement holds also in the odd channel. While it could still in principle happen that the algebraically special pole-skipping points were also common to the two channels, just not related by the Darboux transformations, as we will see in Section~\ref{subsec:PS4_AlgSpec}, this does not occur (except for the special case of the `hydrodynamic' pole-skipping point with $n=0$ for $K=0$).

What we have shown is therefore that the Darboux transformations map even (odd) pole-skipping solutions to odd (even) ones unless the pole-skipping solutions correspond to the algebraically special pole-skipping points.

We now comment also on the cases that fall into the category \ref{outgoingPS}. The situation here is very similar to that in the category \ref{ingoingPS}. Now, however, if $c(\omega,\mu)=0$, then it is the \emph{outgoing} boundary conditions that fails to uniquely determine the solution of the master equation. We will call any such point an {\it (outgoing) pole-skipping point}. In fact, it should be noted that it is generally true that a shift of $\omega_n \to -\omega_n$ changes an ingoing pole-skipping mode in an outgoing one. This property follows directly from Eqs.~\eqref{around_horizon} and \eqref{apparent} and the form of the master equation \eqref{mastereq2}. Unless explicitly stated otherwise, we will always assume all pole-skipping points to be the {\it ingoing} pole-skipping points.

Finally, we summarise here the main three properties of pole-skipping points that have transpired from the master function formalism and Darboux transformations. For completeness, we also include in this summary the special cases with $n=0$ and $n=\pm 1$ that will be discussed below:
\begin{itemize}
    \item Pole-skipping points all have frequencies $\omega = \omega_n = -2 \pi T i n$, with $n\geq -1$. The points with $n\geq 2$ all follow straightforwardly from the master function formalism. Cases with $n=-1$ and $n=1$ require special attention in order to determine what is an ingoing and what an outgoing solution. The case with $n=0$ must be treated independently. 
    
    \item If $(\omega_n,\mu)$ is an ingoing pole-skipping point, then $(-\omega_n,\mu)$ is an outgoing pole-skipping point.
    
    \item If $(\omega_n,\mu)$ is a pole-skipping point in the even (odd) channel, then either
    \begin{itemize}
        \item $\omega^2=\tilde\omega^2$, which means that the pole-skipping point coincides with an algebraically special point. We will call such points \emph{algebraically special pole-skipping points}.
        \item[] or
        \item $\omega^2 \neq \tilde\omega^2$, which means that the pole-skipping point is {\it not} algebraically special. We will refer to such pole-skipping points as the {\it common pole-skipping points} because they simultaneously exist in both the even and the odd channels.
    \end{itemize}
\end{itemize}

\subsection{The ratio of `$B/A$'}\label{sec:BA}

At the pole-skipping points, the ingoing solution contains a free parameter (called $a$) that cannot be fixed in the usual manner. In the vicinity of the pole-skipping point, at $(\omega_0+\delta\omega,\mu_0+\delta \mu)$, however, the solution is nevertheless uniquely determined. The value of $a$ is determined by the slope $\delta\mu/\delta\omega$ with which we choose to approach the pole-skipping point. In other words, the limits of $\omega\to\omega_0$ and $\mu\to\mu_0$ do not commute. 

Consider a solution to the master equation $\psi$, which is ingoing at the horizon and has the following asymptotic form `at infinity': $A\psi^A+B\psi^B$ (cf.~Eq.~\eqref{def:AB}). Due to the dependence of $a$ on $\delta\mu/\delta\omega$, we expect the coefficients $A$ and $B$, which are functions of $\omega$ and $\mu$, to be `undetermined' at the pole-skipping point. Moreover, the ratio $B/A$ that follows from any pole-skipping solution will also be undetermined. Note that in the holographic context, this was discussed extensively in Refs.~\cite{blakeManybodyChaosEnergy2018,blakeHorizonConstraintsHolographic2020}. Nevertheless, the `nature' of this indeterminacy turns out to be such that the quantities $a$, $A$, $B$, $B/A$, and similar ratios, can be written locally as finite, regular fractions. For example, for the ratio of $B/A$, we write $\CA(\omega,\mu) / \CB (\omega,\mu)$, where the two new independent functions $\CA(\omega,\mu)$ and $\CB(\omega,\mu)$ contain no poles. Let $\CA$ and $\CB$ be functions such that $\mathcal A(\omega,\mu)$ and $\mathcal B(\omega,\mu)$ each has an independent `line of zeros' in the $(\omega,\mu)$ space that goes through the point $(\omega_0,\mu_0)$, where `$B/A \sim 0/0$'. We can now expand around this point,

\begin{equation}
\frac{B(\omega_0+\delta\omega,\mu_0+\delta\mu)}{A(\omega_0+\delta\omega,\mu_0+\delta\mu)}= \left.\frac{\partial_\omega \mathcal B(\omega,\mu)+\partial_\mu \mathcal B(\omega,\mu) \frac{\delta\mu}{\delta\omega}}{\partial_\omega \mathcal A(\omega,\mu)+\partial_\mu \mathcal A(\omega,\mu) \frac{\delta\mu}{\delta\omega}}\right|_{(\omega,\mu)=(\omega_0,\mu_0)}, \label{skipping_of_poles}
\end{equation}
having assumed that
\begin{equation}
   \left.\frac{\partial_\omega \mathcal B(\omega,\mu)}{\partial_\mu \mathcal B(\omega,\mu)}
    \right|_{(\omega,\mu)=(\mu_0,\omega_0)} \neq   \left.\frac{\partial_\omega \mathcal A(\omega,\mu)}{\partial_\mu \mathcal A(\omega,\mu)}\right|_{(\omega,\mu)=(\mu_0,\omega_0)}. 
\end{equation}
Note that this assumption restricts us to the cases in which the lines of zeros of $\CA$ and $\CB$ are not tangential to each other at $(\omega_0,\mu_0)$. Eq.~\eqref{skipping_of_poles} explicitly demonstrates how the ratio depends on the slope $\delta\mu/\delta\omega$ at $(\omega_0,\mu_0)$ due to the non-commuting limits of $\omega \to \omega_0$ and $\mu \to \mu_0$. The freedom in the choice of the slope $\delta\mu/\delta\omega$ in Eq.~\eqref{skipping_of_poles} precisely corresponds to the freedom of the choice of $a$ in Eq.~\eqref{free_parameter}. The line of poles is 'skipped' by a zero of $\CB$.

\subsection{Algebraically special pole-skipping points}
\label{subsec:PS4_AlgSpec}

Next, we study explicit details of pole-skipping in the backgrounds of $4d$ black holes. As discussed in Section~\ref{subsec:PS_def_class}, pole-skipping points in 4$d$ can be divided into two broad categories depending on whether they are algebraically special or not. We will first focus on the algebraically special points and show that they can be completely understood analytically. Then, we will explicitly demonstrate that the remaining pole-skipping points are indeed common to both the even and odd channels. 

Pole-skipping points always require us to set $\omega=\omega_n = - 2\pi T i n$. From the point of view of the master fields, finding the corresponding values of $\mu$ then amounts to cancelling the logarithmic terms in the near-horizon expansion. Consider now the pole-skipping modes \eqref{free_parameter} that are also algebraically special and ingoing (cf.~Section~\ref{subsec:Darboux_AlgSpec}), and have $n\geq2$. It is clear that such master fields contain no logarithmic terms as $\Phi(r)$ is an analytic function at $r_0$, and $e^{i\tilde\omega r_*}$ is of the form $(r-r_0)^{i \tilde\omega/4\pi T}\qty(1+c_1 (r-r_0)+\ldots)$. The ingoing algebraically special points are therefore pole-skipping points:
\begin{equation}
    \tilde\omega=\pm \omega_n,~~ n \geq 2, \label{AS0}
\end{equation}
where the $+$ and $-$ signs correspond to the even and odd channels, respectively. Cases with $-1 \leq n \leq 1$ will be discussed below. Eq.~\eqref{AS0} then implies the following relations in the two channels:
\begin{equation}\label{AS1}
\begin{aligned}
    \text{even}:& \qquad \mu \left(\mu-2K\right)=- 3 n \tau ,\\
    \text{odd}:& \qquad \mu \left(\mu-2K\right)=+ 3 n \tau, 
\end{aligned}
\end{equation}
which reveal (two) infinite lines of pole-skipping points per channel. More precisely, in each channel and for every Matsubara level $n \geq 2$ (with the corresponding frequency  $\omega = \omega_n$), we find a pair of pole-skipping points:
\begin{equation}\label{AS2}
\begin{aligned}
&\text{even}:& \qquad \mu_- &= K-\sqrt{K^2 - 3 n \tau}, \\& & \mu_+&=K+\sqrt{K^2 -  3 n \tau},\\
&\text{odd}:& \qquad \mu_- &= K-\sqrt{K^2 + 3 n \tau}, \\& & \mu_+&=K+\sqrt{K^2 + 3 n \tau}. 
\end{aligned}
\end{equation}
While the vales of $\mu_\pm$ are real in the odd channel, they can generally be complex in the even channel. Eq.~\eqref{AS2} therefore provides us with an analytic solution for all algebraically special pole-skipping points with $n\geq 2$.

Next, we analyse the special cases with $n=1$ and $n=-1$. Since algebraically special points have to satisfy the condition $\omega^2 = \tilde\omega^2$, it follows that all such potential pole-skipping points follow from a common analysis. We can choose to set $n=1$ so that $\omega = \omega_1 = - 2 \pi T i$. The discussion then splits into a separate analysis of two cases: $\tilde\omega = \omega_1$ and $\tilde\omega = -\omega_1$:

\begin{itemize}
    \item $\tilde\omega = \omega_1 =-2\pi i T$ \\ In this case, the algebraically special mode is ingoing in the even channel but not in the odd channel. This results in a pair of even channel pole-skipping solutions $\mu_\pm$ that obey Eq.~\eqref{AS2} for $n=1$:
    \begin{align}
        \text{even}: \qquad \mu_-&=K-\sqrt{K^2- 3 \tau}, \\  \mu_+&=K+\sqrt{K^2-3 \tau}. \label{PSEven}
    \end{align}
    This analysis also implies that no pole-skipping point with $n = -1$ exists in the odd channel. 
    
    \item $\tilde\omega = - \omega_1 =2\pi i T$ \\ In this case, the situation is somewhat more involved. We find two solutions for $\mu$, one of which is the solution $\bar\mu$ from Eq.~\eqref{mu0}, only now with the in/outgoing behaviour of the algebraically special solution at the horizon reversed. The second solution does not reverse its in/outgoing character. The two solutions for $\mu$ are therefore
    \begin{align}
        \mu_-&=\bar\mu=K-\sqrt{K^2+3\tau}, \\ \mu_+&=K+\sqrt{K^2+3\tau} ,\label{mus}
    \end{align}
    which we analyse separately. 
    \begin{itemize}
        \item $\mu=K+\sqrt{K^2+3\tau}$: \\
        In this case, the even algebraically special mode is outgoing and the odd solution is ingoing. Hence, this pole-skipping point with $n=1$ exists only in the odd channel.
        
        \item $\mu=\bar\mu=K-\sqrt{K^2+3\tau}$: \\ In this case, the even algebraically special mode is ingoing and the odd solution is outgoing. It would therefore appear that another pole-skipping point with $n=1$ exists in the even channel. To understand the situation better, however, we need to carefully consider the structure of the master function $\psi_+(r)$ defined in Eq.~\eqref{def:master}. Schematically, it behaves as $\psi_+(r)\sim \delta g(r)/\Phi(r)$,        where $\delta g(r)$ is a metric component perturbation from Eq.~\eqref{g_perturbations}. Near the horizon, the ingoing $\psi_+(r)$ scales as $1/\sqrt{r-r_0}$ and  $\Phi(r)$ as $(r-r_0)$. Hence, $\delta g(r)\sim \sqrt{r-r_0}$, which implies that for an ingoing $\psi_+(r)$, we get an \emph{outgoing} $\delta g(r)$. This is therefore not an acceptable pole-skipping solution for $n=1$ from the point of view of metric perturbations. At this point, however, we need to recall the discussion from Section~\ref{subsec:PS_def_class}, which says that for every outgoing pole-skipping solution with frequency $\omega$, we also have an ingoing solution with frequency $-\omega$. Hence, this solution actually corresponds to a pole-skipping solution $\omega = \omega_n$ with $n=-1$. This is the special `chaotic' pole-skipping solution \cite{grozdanovBlackHoleScrambling2018,blakeQuantumHydrodynamicalDescription2017,blakeManybodyChaosEnergy2018,grozdanovConnectionHydrodynamicsQuantum2018}. The reason why it is related to quantum chaos is that $\omega_{-1} = i \lambda_L$, where $\lambda_L$ is the `maximal' Lyapunov exponent $\lambda_L = 2 \pi T$. Moreover, the value of $\mu = \bar \mu$ (introduced in Section~\ref{subsec:Darboux_AlgSpec}) is related to the butterfly velocity associated with the ballistic spread of chaos \cite{grozdanovBlackHoleScrambling2018,blakeQuantumHydrodynamicalDescription2017,blakeManybodyChaosEnergy2018,grozdanovConnectionHydrodynamicsQuantum2018,shenkerBlackHolesButterfly2013,robertsLocalizedShocks2014,shenkerStringyEffectsScrambling2015,maldacenaBoundChaos2015}. We also note that our new, general pole-skipping result at $\mu=\bar\mu$, which holds for any $\Lambda$ and any $K$, agrees with previously known results for the planar black brane \cite{blakeManybodyChaosEnergy2018}, the non-rotating limit of the Kerr black hole \cite{blakeChaosPoleskippingRotating2022} and the hyperbolic black hole \cite{ahnScramblingHyperbolicBlack2019}. 
    \end{itemize}
\end{itemize}
It is important to note that these solutions are special cases of the algebraically special pole-skipping points in Eqs.~\eqref{AS2}. The precise situation will be summarised below. We also note that a careful analysis of the master fields and equations of motion reveals that no other pole-skipping points exist with either $n=1$ or $n=-1$.

Next, we turn our attention to the last special case: pole-skipping points with $\omega = 0$ ($n=0$). The master field formalism is particularly unsuitable to find these points (due to singularities) and special considerations are necessary to uncover them. In this case, it is easiest and most transparent to work directly with the metric perturbations $\delta g_{\mu\nu}$ (as introduced in Section~\ref{subsec:BH_mastereq}). Since these points also turn out to lie on the branches of algebraically special points, we will refer to them by the same name. In order to ascertain whether such a solution corresponds (in the limit of $\omega \to 0$) to an ingoing or an outgoing mode, we need to study the behaviour of the modes in the vicinity of $\omega = 0$. Let us first consider the even channel. There, we can reduce the perturbed Einstein equations to a single second-order equation for the gauge-invariant variable $\bar{w}(r)$ (cf.~Eq.~\eqref{w_def}). In the ingoing Eddington-Finkelstein coordinates, it can be expanded around the horizon as $\bar{w}(r) = \bar{w}_0+\bar{w}_1 (r-r_0)+\ldots$. The Einstein equations give the following relation: 
\begin{equation}
\label{PS_hydro_even}
    \bar{w}_1=\frac{\mu\left(\mu-2K\right)+2 i \omega r_0\left(2K+4\pi T r_0\right)}{r_0^2\left(\mu-2 i \omega r_0\right)\left(4\pi T-2 i \omega\right)}\bar{w}_0.
\end{equation}
We find a pole-skipping solution when the numerator and the denominator on the right-hand-side simultaneously vanish, which for $\omega=0$ occurs when
\begin{equation}
\mu = 0.
\end{equation}
In the odd channel, the analysis proceeds in an analogous manner. We expand $\bar{v}_v(r)= \bar{v}_{0v} + \bar{v}_{1v} (r-r_0)\ldots$, cf.~Eq.~\eqref{v_def}, and use the Einstein equations to find 
\begin{equation}
\label{PS_hydro_odd}
    \bar{v}_{1v}=-\frac{(\mu-2K-2i\omega r_0)(4\pi T-i\omega)}{i\omega r_0^2(4\pi T-2i\omega)}\bar{v}_{0v}.
\end{equation}
The numerator and the denominator of the right-hand-side now vanish for $\omega = 0$ when
\begin{equation}
    \mu = 2 K.
\end{equation}
These are the only two pole-skipping points with $\omega = 0$ --- one in the even and one in the odd channel. The solutions are again special cases of a single branch of the algebraically special solutions $\omega(\mu)$ in Eq.~\eqref{AS2}. Note also that for the case of a black brane (in asymptotically AdS spacetime), both channels have a pole-skipping point at $\omega = 0$ and $\mu = 0$, which corresponds to the pole-skipping of the hydrodynamic gapless sound and diffusive modes in the even and odd channels, respectively \cite{grozdanovComplexLifeHydrodynamic2019}. We note, moreover, that the above expressions also yield the pole-skipping points with $n = \pm 1$ in both channels. 

In summary, in the even channel, the algebraically special pole-skipping points with $\omega = \omega_n$ exist for $n \geq - 1$. The solutions have the following values of $\mu$:
\begin{equation}\label{even_ASPS}
\begin{aligned}
    n=-1:& & &\mu = K - \sqrt{K^2 + 3\tau},   \\
    n=0:& & &\mu = 0, \\
    n \geq 1:& & &\mu = K \pm \sqrt{K^2 - 3 n \tau}. 
\end{aligned}
\end{equation}
Only a single solution exists for each of $n=-1$ and $n=0$, while for $n\geq 1$, we have a pair of solutions for every $n$. In the odd channel, algebraically special pole-skipping solutions exist for $n \geq 0$ and have the following values of $\mu$:
\begin{equation}\label{odd_ASPS}
\begin{aligned}
    n=0:& & &\mu = 2 K, \\
    n=1:& & &\mu = K + \sqrt{K^2 + 3\tau}, \\
    n \geq 2:& & &\mu = K \pm \sqrt{K^2 + 3 n \tau}. 
\end{aligned}
\end{equation}
Here, only a single solution exists for each of $n=0$ and $n=1$, while for $n\geq 2$, we have a pair of solutions for every $n$. Finally, it is also important to note that the `chaotic' pole-skipping solution with $n=-1$ in the even channel and the $n=1$ solution in the odd channel break the symmetry of Darboux transformations between the even and odd channels. 

\subsection{Common pole-skipping points}
\label{subsec:PS4_systematic}

In this section, we investigate the remaining infinite set of pole-skipping points and explicitly verify that, as argued in Section~\ref{subsec:PS_def_class}, they are indeed common to both channel. The simplest way to find these points, which need to be analysed numerically, is by using the `determinant method' first described in Ref.~\cite{blakeHorizonConstraintsHolographic2020} (see also Ref.~\cite{ahnClassifyingPoleskippingPoints2021}). As can be checked explicitly, for example by analysing pole-skipping directly at the level of metric perturbations, all pole-skipping points in black hole backgrounds considered here with $-1 \leq n \leq 2$ are algebraically special. Hence, to look for the common points (all with $n\geq 3$), we can employ the determinant method directly by using the master field variables $\psi_\pm$. 

The analysis is most conveniently performed in the ingoing Eddington-Finkelstein (EF) coordinates with $v=t+r_*$. To write the master field equations, we relate $\psi_\pm$ expressed in the`Schwarzschild' coordinates (cf.~Eq.~\eqref{fourier}) to those in the EF coordinates as
\begin{equation}
    \psi_\text{EF}(r)=\psi(r)e^{i\omega r_*}.
\end{equation}
The master equation \eqref{mastereq2} then takes the following form in both channels:
\begin{equation}
    f(r)\psi''_\text{EF}(r)+\left(f'(r)-2i\omega\right)\psi'_\text{EF}(r)-V(r)\psi_\text{EF}(r)=0. \label{mastereqEF}
\end{equation}
In the EF coordinates, the ingoing pole-skipping solutions arise when we find two analytic (regular) functions at the horizon. We first expand all functions that appear in the master equations:
\begin{align}
f(r)&=\sum_{n=1}^\infty (r-r_0)^n f_n, \\
V(r)&=\sum_{n=0}^\infty (r-r_0)^n V_n, \\
\psi_\text{EF}(r)&=\sum_{n=0}^\infty (r-r_0)^n \psi_n,
\end{align}
where $f_1=4\pi T$. Eq.~\eqref{mastereqEF} then gives following system of equations:

\begin{equation}
\begin{pmatrix}
-V_0 & f_1-2i\omega & 0 & 0 & 0 & \cdots \\ 
-V_1 & 2f_2-V_0 & 2(2f_1-2i\omega) & 0 & 0 & \cdots \\ 
-V_2 & 3f_3-V_1 & 6f_2-V_0 & 3(3f_1-2i\omega) & 0  & \cdots \\ 
-V_3 & 4 f_4-V_2 & 8f_3-V_1 & 12f_2-V_0 & 4(4f_1-2i\omega)  & \cdots \\ 
-V_4 & 5 f_5-V_3 & 10f_4-V_2 & 15f_3-V_1 & 20f_2-V_0  & \cdots \\ 
\vdots & \vdots & \vdots & \vdots & \vdots  & \ddots \\ 
\end{pmatrix} 
\begin{pmatrix}
\psi_0 \\ \psi_1 \\ \psi_2 \\ \psi_3 \\ \psi_4 \\ \vdots
\end{pmatrix}=\begin{pmatrix}
0 \\ 0 \\ 0 \\ 0 \\ 0 \\ \vdots  \label{the_matrix}
\end{pmatrix}.
\end{equation}

We can now use the matrix in the above expression to find the pole-skipping points by the following procedure \cite{blakeHorizonConstraintsHolographic2020}:
\begin{enumerate}
    \item Set $\omega=\omega_n$. 
    \item Calculate the determinant $D_n(\omega_n,\mu)$ of the upper-left $n \times n$ matrix.
    \item Solve $D_n(\omega_n,\mu)=0$.
\end{enumerate}
The determinant equation is then solved for increasing values of the integer $n$. It is important to note that when analysing a particular channel, this procedure yields both the algebraically special and the common pole-skipping points, simultaneously. At present, unlike for the algebraically special modes, the solutions to $D_n(\omega_n,\mu)=0$ cannot be found in analytically in closed form, which in turn means that no analytic solution exists for the common points. The fact that beyond the algebraically special points, all other pole-skipping points are shared between the two channels (which we argued must be true by using the Darboux transformations in Section~\ref{subsec:PS_def_class}) is therefore observed and verified numerically for specific values of $n$.

Beyond giving us the values of $\omega = \omega_n$ and $\mu$ at the pole-skipping points, the determinant also allows us to elaborate on the relation between pole-skipping and the slope of $\delta \mu / \delta \omega$. In particular, we start by deriving the relation between $\psi_n$ and $\psi_0$:
\begin{equation}
    \frac{D_n(\omega,\mu)}{N_n(\omega)}\psi_0-\left(2i\omega-n f_1\right) \psi_n=0, \label{psi_n}
\end{equation}
where
\begin{equation}
    N_n(\omega)=(n-1)! \left(2i\omega-f_1\right) \left(2i\omega-2f_1\right)\ldots  \left(2i\omega-(n-1)f_1\right),
\end{equation}
which holds for any $n \geq 1$. All further coefficients, $\psi_{n+1}$, $\psi_{n+2}$, $\ldots$, are then calculated recursively. For pole-skipping with $\omega=\omega_n$ and a corresponding $\mu=\mu_n$, where $D_n(\mu_n,\omega_n)=0$, Eq.~\eqref{psi_n} does not impose a condition on $\psi_n$, leaving us with a free parameter. Expanding Eq.~\eqref{psi_n} around this pole-skipping point then yields

\begin{align}
    \left. \qty[
    \frac{K_n(\omega,\mu)}{N_n(\omega)}
    \psi_0-2i \psi_n] \right|_{(\omega,\mu)=(\omega_n,\mu_n)}=0, \label{nth_term} \\
    K_n(\omega,\mu)=\partial_\mu D_n(\omega,\mu)\frac{\delta\mu}{\delta\omega} +\partial_\omega D_n(\omega,\mu).
\end{align}

We see that the ratio of $\psi_n$ and $\psi_0$ explicitly depends on the slope $\delta\mu/ \delta\omega$, as is usual with the pole-skipping points. It should be noted that if, furthermore, $\partial_\mu D_n(\omega,\mu)=0$, the free parameter in the solution is lost. Such pole-skipping points, which do not exhibit standard pole-skipping properties, are called anomalous \cite{blakeHorizonConstraintsHolographic2020} or type II \cite{ahnClassifyingPoleskippingPoints2021}. We will encounter them in Section~\ref{subsec:PS4_cosmological}.

\begin{figure*}[ht!]
     \begin{subfigure}[b]{0.325\textwidth}
         \centering
         \includegraphics[width=\textwidth]{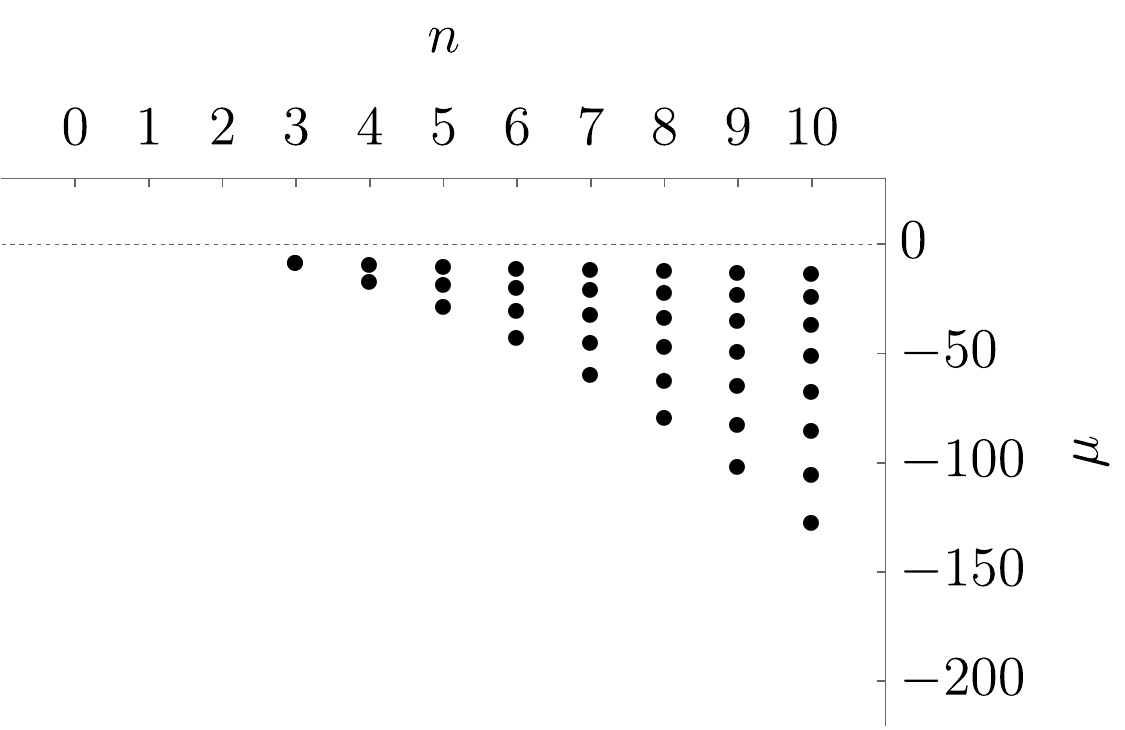}
         \subcaption{$K=1,\tau=\frac{1}{2}$}
     \end{subfigure}
     \begin{subfigure}[b]{0.325\textwidth}
         \centering
         \includegraphics[width=\textwidth]{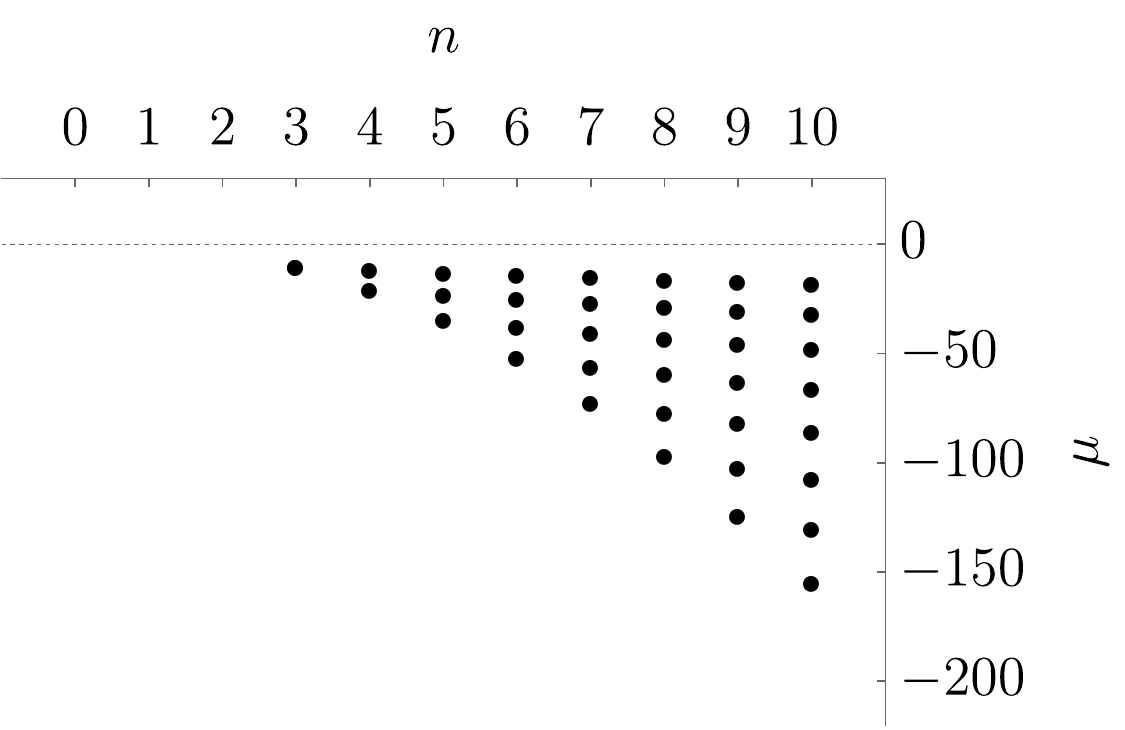}
                  \subcaption{$K=1,\tau=1$}
     \end{subfigure}
     \begin{subfigure}[b]{0.325\textwidth}
         \centering
         \includegraphics[width=\textwidth]{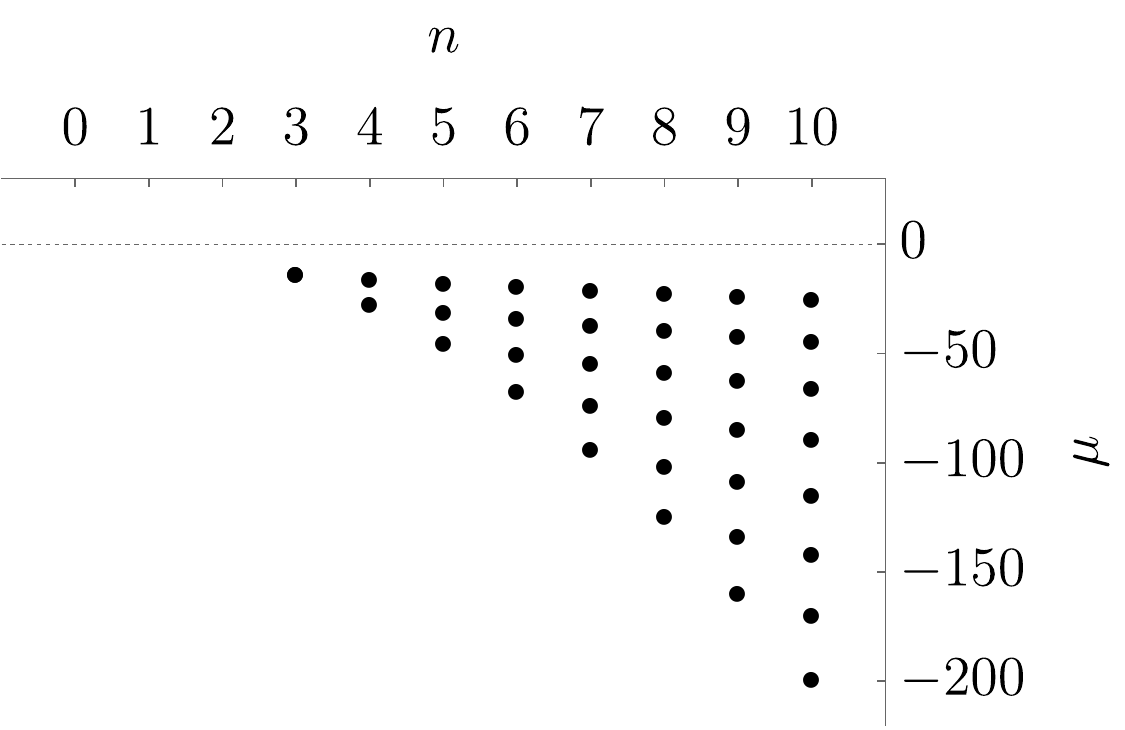}
                  \subcaption{$K=1,\tau=2$}
     \end{subfigure}
     \hfill
        \caption{Locations of the common pole-skipping points (black dots) along the real $\mu$ axis for spherical black holes ($K=1$) at different Matsubara levels $n$. Three panels correspond to three choices of the parameter $\tau$, giving geometries that are asymptotically dS, flat and AdS (respectively, from left to right).}
        \label{fig:common1}
\end{figure*}

Finally, we can now summarise the known results for all pole-skipping points, where the explicit expressions correspond to the algebraically special points found in Section~\ref{subsec:PS4_AlgSpec} (cf.~Eqs.~\eqref{even_ASPS} and \eqref{odd_ASPS}). As for the common points discussed in this section, we only state that at level $n$, there exist $n-2$ such points (as can be seen for example in Figure~\ref{fig:common1}). This fact can also be ascertained from the polynomial order of the equations $D_n(\omega_n,\mu)=0$ at different $n$. The results are: \\
\begin{center}
\begin{tabular}{ |c|c c| } 
 \hline
  $n$       & even channel                  & odd channel               \\ \hline
 $-1$       & $K-\sqrt{K^2+3\tau}$      &  $\times$                      \\ \hline
 $0$        & $0$                       & $2K$                  \\ \hline
 $1$        & $K+\sqrt{K^2-3\tau}$      & $K+\sqrt{K^2+3\tau}$  \\ 
            & $K-\sqrt{K^2-3\tau}$      &  $\times$                      \\ \hline
 $\geq 2$   & $K+\sqrt{K^2-3n\tau}$     & $K+\sqrt{K^2+3n\tau}$ \\ 
            & $K-\sqrt{K^2-3n\tau}$     & $K-\sqrt{K^2+3n\tau}$ \\
            & \multicolumn{2}{|c|}{$n-2$ common pole-skipping} \\
            & \multicolumn{2}{|c|}{points with $\mu<0$} \\
            \hline
\end{tabular}
\end{center}

\vspace{14pt}
It is also worth stating the following general facts about pole-skipping in our examples. For black branes ($K=0$), there exists a scaling transformation that leaves the master equations invariant up to a multiplicative constant, namely, $\tau\rightarrow\alpha^2\tau$, $\omega\rightarrow\alpha^2\omega$, $\mu\rightarrow\alpha\mu$, $r\rightarrow r/\alpha$, with $\alpha$ a constant. This implies that the knowledge of pole-skipping points at any specific $\tau$ allows us to easily find the pole-skipping points for any $\tau$ via the above rescaling. For spherical and hyperbolic black holes ($K = \pm 1$), the situation is more involved. Nevertheless, we observe a symmetry between the pole-skipping points with $K=1$ and those with $K=-1$. Here, we only state without proof that given a pole-skipping point $(\omega_n, \mu_n)$ with $K=1$, there exists a pole-skipping point with $K=-1$ at $(\omega_n,\mu_n-2)$. This statement can be easily checked numerically. Moreover, in the limit of large $\mu$ and large $\tau$, the effect of the horizon topology encoded in $K$ becomes negligible so that all three cases exhibit similar large-parameter scaling properties to the pole-skipping points of a black brane ($K=0$). More specifically, pole-skipping values of $\mu$ will scale as $\mu_n \propto \sqrt{\tau}$, or equivalently, $\mu_n \propto \sqrt[3]{-M^2\Lambda}$, for $\mu_n \gg 1$ and $\tau \gg 1$ (or, equivalently, for $M^2 \Lambda \ll -1$).

\begin{figure*}[ht!]
     \begin{subfigure}[b]{0.325\textwidth}
         \centering
         \includegraphics[width=\textwidth]{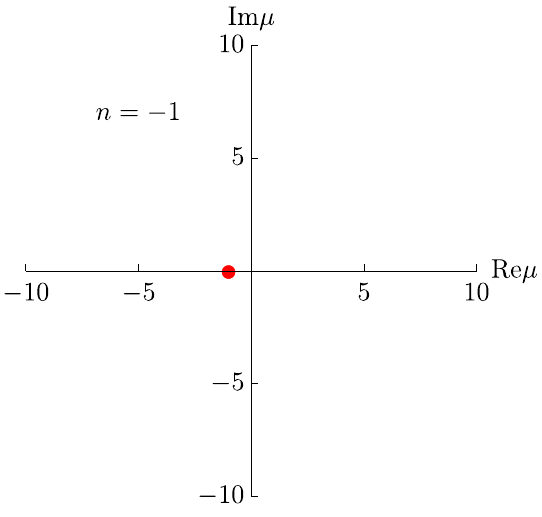}
     \end{subfigure}
     \begin{subfigure}[b]{0.325\textwidth}
         \centering
         \includegraphics[width=\textwidth]{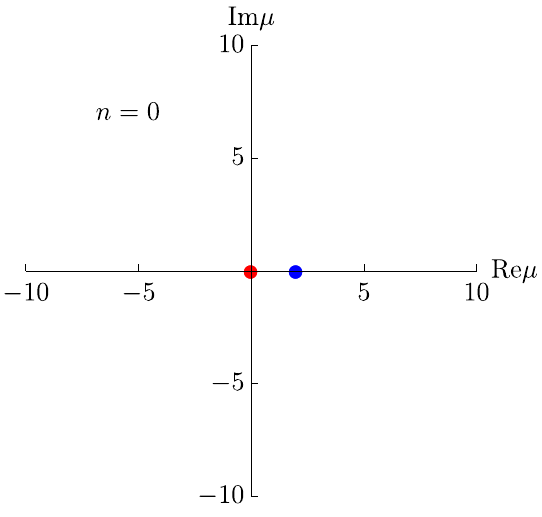}
     \end{subfigure}
          \begin{subfigure}[b]{0.325\textwidth}
         \centering
         \includegraphics[width=\textwidth]{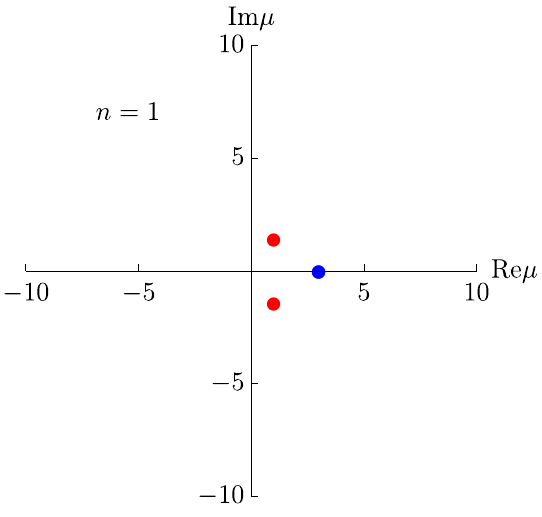}
     \end{subfigure}
          \begin{subfigure}[b]{0.325\textwidth}
         \centering
         \includegraphics[width=\textwidth]{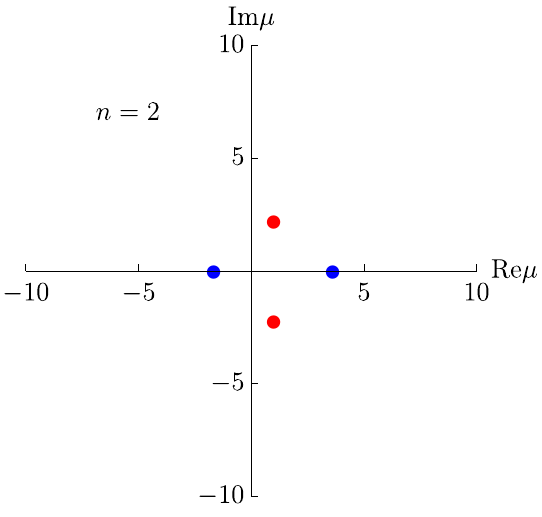}
     \end{subfigure}
    \begin{subfigure}[b]{0.325\textwidth}
         \centering
         \includegraphics[width=\textwidth]{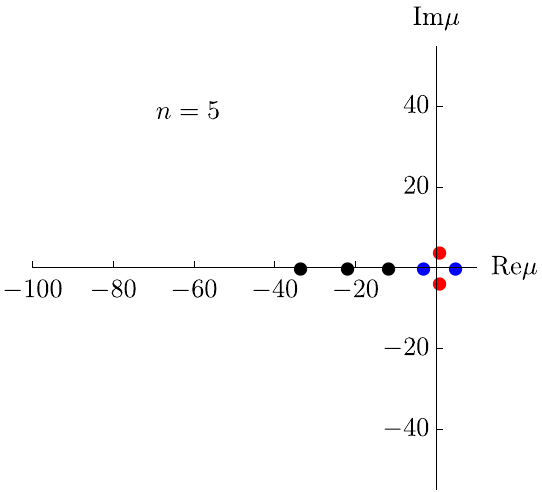}
     \end{subfigure}
     \begin{subfigure}[b]{0.325\textwidth}
         \centering
         \includegraphics[width=\textwidth]{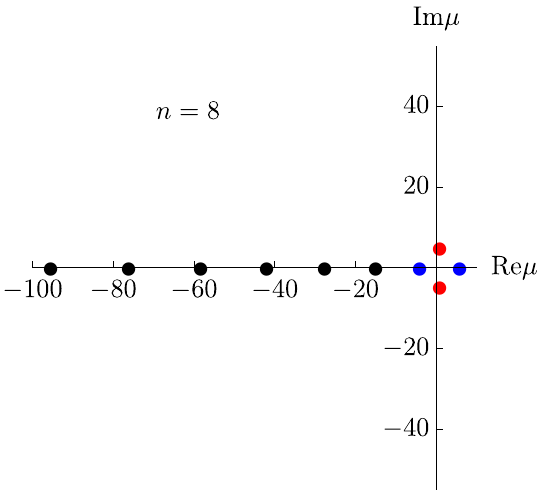}
     \end{subfigure}

        \caption{All pole-skipping points plotted in the complex $\mu$ plane for six choices of the Matsubara level $n$ of the asymptotically flat spherical Schwarzschild black hole with $K=1$ and $\tau = 1$. Red and blue dots depict the algebraically special solutions in the even and odd channels, respectively. Black dots are the common points shared between the two channels.}
        \label{fig:pole-skips}
\end{figure*}

We show three examples of the distribution of the common pole-skipping points for spherical black holes at different Matsubara levels $n$ in Figure~\ref{fig:common1}. All those points have $\mu \in \mathbb{R}$. Then, in Figure~\ref{fig:pole-skips}, we plot an example of the distribution of all pole-skipping points in the complex $\mu$ plane for the asymptotically flat and spherical Schwarzschild black hole ($K=1$ and $\tau = 1$).

\subsection{More on the algebraically special pole-skipping points}
\label{subsec:revisited}

Using the results from Section~\ref{subsec:PS4_systematic}, we can further elaborate on the characteristics of the algebraically special pole-skipping points. For simplicity (the ability to directly use the master field language), we will only focus on points with $n\geq1$. Recall, however, that the $n=1$ master field analysis also reveals the `chaotic' pole-skipping point with $n=-1$ in the even channel. An algebraically special pole-skipping point always yields a master field solution of the form 
\begin{equation}
    \psi_\text{EF}(r)=\tilde\psi_\text{EF}(r) + a \tilde\chi_\text{EF}(r).
\end{equation}
Note that, hereon, we will suppress the subscripts indicating our use of the ingoing EF coordinates. As already discussed, the parameter $a$ is the free parameter that depends on the slope $\delta \mu/\delta \omega$. The goal of this section is to explicitly derive this dependence, i.e., the function $a(\delta \mu/\delta \omega)$.

This can be done by using the expression in Eq.~\eqref{nth_term}. Since, by construction, $\tilde\chi_\pm(r)$ will not have a constant term at the horizon, we can write
\begin{equation}
    \psi_0=\tilde\psi_0.
\end{equation}
We can then set $\omega$ and $\mu$ to the pole-skipping values $\omega=\omega_n$ and $\mu=\mu_n$, and consider the $n$-th order term in the expansion around the horizon, $\psi_n=\tilde\psi_n+a \tilde\chi_n$. Using the explicit solutions of the algebraically special modes from Eqs.~\eqref{def:AS} and \eqref{chis} allows us to first write
\begin{equation}
    \tilde\chi^{\pm}_n=\frac{1}{ 4 \pi n T \tilde\psi^{\pm}_0}, \label{chi0}
\end{equation}
and then arrive at 
\begin{equation}
        a_\pm= 4\pi n T \tilde\psi^{\pm}_0 \left. \qty[
    \frac{K_n(\omega,\mu)}{2 i N_n(\omega)}
    \tilde\psi^{\pm}_0-\tilde\psi^{\pm}_n] \right|_{(\omega,\mu)=(\omega_n,\mu_n)}. \label{a_equation}
\end{equation}
This expression can be simplified by using the fact that $a_\pm = 0 $ when $\delta \omega/\delta \mu=\pm \dd \tilde\omega/\dd \mu$. Hence, in the two channels, we finally obtain the sought relation
\begin{equation}
     a_\pm= 4\pi n T \tilde\psi^{\pm}_0 \qty(\tilde\psi^{\pm}_n-\frac{\partial_\omega D_n(\omega,\mu)}{2iN_n(\omega)}\tilde\psi^{\pm}_0) \left.\qty(\pm\frac{\dd \tilde\omega}{\dd\mu}\frac{\delta\mu}{\delta\omega}-1)\right|_{(\omega,\mu)=(\omega_n,\mu_n)}. \label{new_a}
\end{equation}

For a black hole with the spherical horizon ($K=1$) in the asymptotically flat space ($\Lambda=0$) --- i.e., the Schwarzschlid black hole --- the above expression gives a simple relation between $a_\pm$ and the asymptotic ratio $B/A$. In particular, the exponential asymptotic behaviour of the master functions allows us to use the integrals in Eq.~\eqref{chis} to argue that given an ingoing $\tilde\psi_\pm$ at the horizon, $\tilde\chi_\pm$ will be asymptotically outgoing (in addition to being outgoing at the horizon by construction). It then immediately follows that
\begin{equation}
\frac{B_\pm}{A_\pm} \propto a_\pm,
\end{equation}
where the proportionality factor is independent of $\delta\mu/\delta\omega$. We can also conclude by using Eq.~\eqref{skipping_of_poles} that for the algebraically special pole-skipping points, the line of poles points in the direction of $\delta \omega=0$ (i.e., $\delta\mu/\delta\omega=\infty$) while the line of zeros points in the direction of $\delta\omega/\delta\mu=\pm\dd \tilde\omega/\dd \mu$. Moreover, by setting $\delta \mu = 0$, which is used when calculating for example the time evolution of gravitational perturbations, we still retain a non-zero $a$. This means that, unlike what was previously argued for in Ref.~\cite{anderssonTotalTransmissionSchwarzschild1994}, algebraically special points that are also pole-skipping points are not totally transmissive in the appropriate limit.

\section{Special cases of pole-skipping in $4d$ black hole backgrounds}
\label{sec:PS4}

In this section, we study three explicit examples of pole-skipping in 4$d$ spacetimes. The first two are special cases of the analysis in Section~\ref{sec:PS}. Namely, we consider spherical black holes and a black brane in an asymptotically AdS space (which is particularly relevant to the AdS${}_4$/CFT${}_3$ holography). At the end, we also investigate certain details of pole-skipping at the cosmological horizon in asymptotically dS spaces.  

\subsection{Spherical black holes}
\label{subsec:PS4_spherical}

We start by studying the spherically symmetric (Schwarzschild) black holes with $K=1$, but with any $\Lambda$. So far, our treatment of the parameter $\mu \in \mathbb{C}$ has been kept completely general. However, for a `physical' gravitational wave propagating in the spherically symmetric black hole spacetime, $\mu = \ell (\ell+1)$ (cf.~Eq.~\eqref{mu-l-forK1}), where $\ell$ is an integer and $\ell \geq 0$. Note, however, that $\ell = 0$ and $\ell = 1$ modes are static perturbations and the modes with $\ell \geq 2$ are dynamical gravitational waves. Hence, $\mu$ is also an integer with $\mu = 0$ and $\mu = 2$ that correspond to the static perturbations with $\ell = 0$ and $\ell = 1$, respectively, and $\mu \geq 6$ for modes with $\ell \geq 2$. Recall now from Section~\ref{subsec:PS4_systematic} that the pole-skipping points that are {\it not} algebraically special have a negative $\mu$. Hence, if we restrict our attention to only the `physical' gravitational waves (not their analytic continuations or the complex structure of Green's functions), this fact makes the analysis of such pole-skipping dramatically simpler for spherical black holes, which are certainly of greatest interest for astrophysics. 

The algebraically special pole-skipping points  are given by Eqs.~\eqref{even_ASPS} and \eqref{odd_ASPS} with $K=1$. In particular, in the even channel, we have
\begin{equation}\label{evenk1}
\begin{aligned}
    n=-1:& & &\mu = 1 - \sqrt{1 + 3\tau},   \\
    n=0:& & &\mu = 0, \\
    n \geq 1:& & &\mu = 1 \pm \sqrt{1 - 3 n \tau}.
\end{aligned}
\end{equation}
In the odd channel, we have
\begin{equation}\label{oddk1}
\begin{aligned}
    n=0:& & &\mu = 2, \\
    n=1:& & &\mu = 1 + \sqrt{1 + 3\tau}, \\
    n \geq 2:& & &\mu = 1 \pm \sqrt{1 + 3 n \tau}. 
\end{aligned}
\end{equation}
Recall that $\tau >0$ and that for `physical' pole-skipping points, we can only have $\mu = \{0, 2, 6, 12, \ldots\}$. Note also that for the sets of double branches (for $n\geq 1$ in the even and for $n\geq 2$ in the odd channels), it is useful to recall Eq.~\eqref{AS1} for $K=1$: 
\begin{equation}
    \mu \left(\mu-2\right)=\mp 3 n \tau, \label{AS_spherical}
\end{equation}
where  $-$ sign corresponds to the even and the $+$ sign in the odd channel.

By taking into account the possible ranges of parameters, it is easy to see that in the even channel, we can only have a single such solution (cf.~Eq.~\eqref{evenk1}):
\begin{equation}
(\ell, n) = \{(0,0)\},
\end{equation}
which is a static mode. 

On the other hand, in the odd channel, an infinite set of `physical' pole-skipping points exits. Since $\mu \geq 0$, the only relevant branch of odd modes is (cf.~Eq.~\eqref{oddk1})
\begin{equation}
    \mu = \ell \left(\ell + 1\right) = 1 + \sqrt{1 + 3 n \tau}, ~~ n \geq 0.
\end{equation}
The equation has no solution for $\ell=0$ but it can in principle (depending on the value of $\tau$) be solved for other integer value of $\ell \geq 1$. The resulting relation between $n$ and $\ell$ is  
\begin{equation}\label{n_l_relation}
    n=\frac{\left(\ell-1\right) \ell \left(\ell+1\right) \left(\ell+2\right)}{3\tau}.
\end{equation}

When is this equation satisfied for integer values of $n \geq 0 $ and $\ell \geq 1$? For asymptotically flat and spherical Schwarzschild black holes, $\tau = 1$ (recall the definition from Eqs.~\eqref{def:tau}--\eqref{taulambda}). Since the numerator of \eqref{n_l_relation} is a product of four consecutive integers, which is always divisible by $3$, it is clear that an integer $n$ exists for all $\ell = 1,2,3,\ldots$ such that Eq.~\eqref{n_l_relation} is satisfied. In other words, for asymptotically flat black holes, we have one pole-skipping point in the odd channel of perturbations for every `physical' $\ell$. The first few points are labeled by the following integers: 
\begin{equation}
    (\ell, n) = \{ (1,0), (2,8),(3,40),(4,120),(5,280),\ldots \}. 
\end{equation}
Since the solutions are spherical harmonics, it is worth keeping in mind that due to the degeneracy of these solutions (for $m\in \mathbb{Z}$ with $-\ell \leq m \leq \ell$) each of the above pole-skipping points amounts to $2\ell+1$ solutions. Note also that the first solution in the set with $\ell = 1$ and $n=0$ is a static mode.

We now also consider asymptotically AdS and dS spaces. In such cases, $\tau$ can take a continuum of values. (Recall the plot from Figure~\ref{fig:f}.) What we find from Eq.~\eqref{n_l_relation} is that in asymptotically dS spaces ($0 < \tau < 1$), `physical' pole-skipping modes exist with integer $n$ for every integer $\ell \geq 1$ if either $\tau = 1/3$ or $\tau = 2/3$. In asymptotically AdS spaces, this is possible if $\tau = 4/3$. More generally, if $\tau$ is a rational number, we still obtain an infinite number of pole-skipping points with integer $n$ but not necessarily for every $\ell \geq 1$. On the other hand, if $\tau$ is an irrational number, no `physical' pole-skipping points can be found beyond the static solution with $n=0$. 

The main effect of pole-skipping is in the fact that the ratio of $B/A$, which characterises the gravitational waves at asymptotic infinity, is not uniquely determined and depends on the slope $\delta \mu /\delta \omega$ (cf.~Section~\ref{sec:BA}). The limits of $\mu$ and $\omega$ approaching the pole-skipping point do not commute. If, however, we restrict ourselves to integer values of $\mu$, then one may justifiably doubt that such limits are of any relevance to the behaviour of physical gravitational waves in the backgrounds of spherically symmetric black holes. For example, if one considers the real-time propagation of such modes, then the transformation to real time and space involves a sum over $\ell$ and an integral over $\omega$. Hence, for such problems, $\ell$ and therefore $\mu$ naturally take discrete values.

\subsection{Asymptotically AdS black brane and the dual hydrodynamic mode}

A 4$d$ black hole with a translationally invariant horizon ($K=0$) in an asymptotically AdS space ($\Lambda < 0$)  --- a black brane --- is holographically dual to a thermal state in a 3$d$ conformal field theory \cite{zaanenHolographicDualityCondensed2015,ammonGaugeGravityDuality2015,hartnollHolographicQuantumMatter2016}. In terms of its gravitational perturbations, its lowest, gapless quasinormal mode is dual to a hydrodynamic mode, i.e., $\omega(k\to 0 ) = 0$. In the odd channel, this is a diffusive mode and in the even channel, this is a pair of sound modes. As was shown in Ref.~\cite{grozdanovBoundsTransportUnivalence2020}, in the odd channel, the `diffusive' QNM passes through an infinite sequence of pole-skipping points $\omega = \omega_n = - 2\pi T i n$ at the corresponding values of wavevectors $k = k_n$, which can be computed analytically:
\begin{equation}\label{hydroPS}
    k_n = \frac{4\pi T}{\sqrt{3}} n^{1/4}, 
\end{equation}
for $n = 0,1,2,\ldots$.

It is easy to check that Eq.~\eqref{hydroPS} is a special case of our general result from Eq.~\eqref{odd_ASPS}. In particular, using the fact that for a black brane, $K=0$, $\mu = k^2$ (cf.~Eq.~\eqref{mu-k-forK0}) and taking $\Lambda = -3$ (in units of the `AdS radius' set to one), then the defining equation for $\tau$ (cf.~Eq.~\eqref{def:tau}) gives
\begin{equation}
    \tau = \frac{(4 \pi T)^4}{3^3}.
\end{equation}
Finally, using Eq.~\eqref{odd_ASPS} for the odd channel with $k^2$ real and non-negative, i.e., $k^2 = \sqrt{3 n \tau}$, immediately yields the holographic hydrodynamic pole-skipping result Eq.~\eqref{hydroPS} as its real and non-negative solution for $k = k_n$.

\subsection{The cosmological horizon in asymptotically dS spaces}
\label{subsec:PS4_cosmological}

In asymptotically dS spaces ($\Lambda > 0$), only black holes with spherical horizons ($K=1$) can exist. In such spaces, the event horizon of a black hole is not the only place where one can expect pole-skipping to occur. The second horizon, also a zero of $f(r)$, is the {\it cosmological horizon} at $r = r_c$ (cf.~Section~\ref{subsec:BH_background}). Just like at $r_0$, at $r_c$ too, gravitational pertubation equations have a regular singular point. The only difference now is that we want the modes at $r_c$ to be outgoing --- they all `leave the universe'. Beyond that, the entire discussion of pole-skipping at the cosmological horizon follows our analysis from previous sections. 

Before starting with the analysis, it is useful to note that for the cosmological horizon, $f'(r_c) < 0$. Thus, if the definition of the temperature \eqref{def:temperature} were to be retained, this would imply that $T < 0$. For present purposes, it suffices to formally keep this definition of the {\it symbol} $T$ (which is not the thermodynamic temperature). For a sensible definition of the Hawking temperature and other thermodynamic properties of de Sitter space, see for example Ref.~\cite{spradlinHouchesLecturesSitter2001}. In terms of $\tau$, we now have $\tau \in [-1/3,0)$ (see Figure~\ref{fig:f}). Note that $\tau$ is now not a bijective function of $4M^2\Lambda$ (the parameter used in Figure~\ref{fig:f}). We also note that since we will be looking for outgoing pole-skipping points, we expect from general arguments that we will find the resulting frequencies to be $\omega_n = - 2\pi T i n $ with $n \leq 1$ instead of $n \geq -1$ (recall our discussion in Section~\ref{subsec:PS_def_class}). Since $T < 0$, this, however, means that $\omega_n$ do again lie along the negative (or more precisely, non-positive) imaginary axis for $n \leq 0$, just as for pole-skipping at the event horizon. 

\begin{figure*}[h!]
\centering
    \begin{subfigure}[b]{0.325\textwidth}
         \centering
         \includegraphics[width=\textwidth]{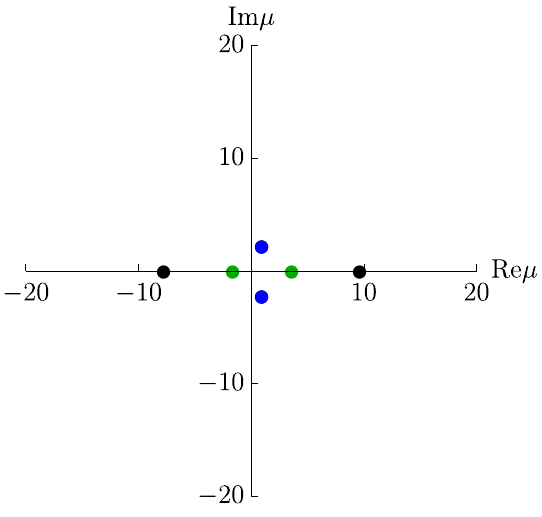}
         \subcaption{$n=-6, \tau=-\frac{1}{3}, r_c=6M$}
     \end{subfigure}
     \begin{subfigure}[b]{0.325\textwidth}
         \centering
         \includegraphics[width=\textwidth]{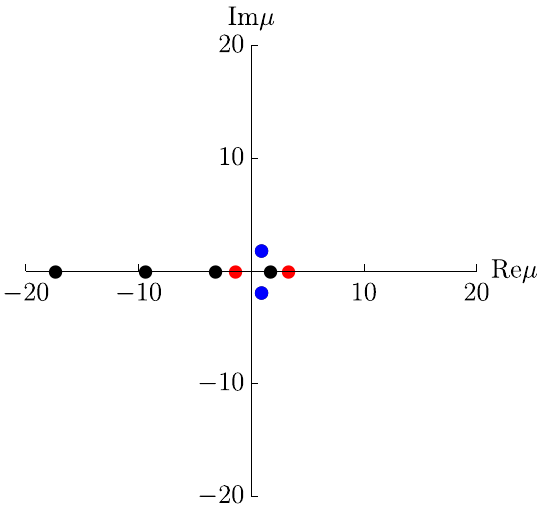}
         \subcaption{$n=-6, \tau=-\frac{1}{4}, r_c=4M$}
     \end{subfigure}
     \begin{subfigure}[b]{0.325\textwidth}
         \centering
         \includegraphics[width=\textwidth]{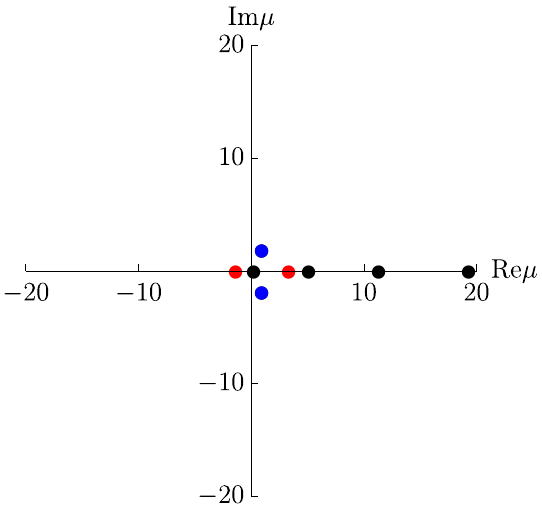}
         \subcaption{$n=-6, \tau=-\frac{1}{4}, r_c=12M$}
     \end{subfigure}
        \caption{The cosmological horizon pole-skipping points in the complex $\mu$ plane plotted for $n=-6$ and for three different positions of the cosmological horizon. Red and blue dots depict the algebraically special solutions in the even and odd channels, respectively. Black dots are the common points shared between the two channels. In the even channel, the algebraically special points can overlap with common points, yielding anomalous pole-skipping points, which are plotted with green dots. For a given $\tau$, the solutions in the two channels can be related by taking $\mu\rightarrow 1-\mu$. Moreover, note that for the `extreme' value of $\tau=-1/3$, the spectrum is symmetric across the line of  $\text{Re}\mu=1$.}
        \label{fig:pole-skips_dS}
\end{figure*}

We can now repeat the pole-skipping analysis discussed in previous sections, splitting the points again into a set of algebraically special and common pole-skipping points. For the first set, we now find that in the even channel,
\begin{equation}\label{even_dS}
\begin{aligned}
    n=1:& & &\mu = 1 - \sqrt{1 + 3\tau},   \\
    n=0:& & &\mu = 0, \\
    n \leq -1:& & &\mu = 1 \pm \sqrt{1 + 3 n \tau},
\end{aligned}
\end{equation}
and in the odd channel, 
\begin{equation}\label{odd_dS}
\begin{aligned}
    n=0:& & &\mu = 2 , \\
    n=-1:& & &\mu = 1 + \sqrt{1 + 3\tau}, \\
    n \leq -2:& & &\mu = 1 \pm \sqrt{1 - 3 n \tau}. 
\end{aligned}
\end{equation}
As for the common pole-skipping points, we now have $-n-2$ solutions, all with real $\mu$. Hence, we find non-trivial common pole-skipping points for all Matsubara frequencies with $n \leq -3$. In Figure~\ref{fig:pole-skips_dS}, we show the values of the pole-skipping points in the even and odd channels for three choices of $r_c$ and $\tau$ at $n=-6$. It is worth noting that the algebraically special points in the even channel have real $\mu$, while in the odd channel they have complex $\mu$. In the even channel, we also find anomalous pole-skipping points.

Next, we turn our attention to the empty de Sitter space in which only the cosmological horizon exists.  Now, the even and the odd potentials are the same:
\begin{equation}
    V_+(r)=V_-(r)=\frac{\ell\left(\ell+1\right)}{r^2},
\end{equation}
with the emblackening factor reducing to a simple `BTZ-like' quadratic form:
\begin{equation}
    f(r)=1-\frac{r^2}{r_c^2}.
\end{equation}
It follows that we have $T=-1/2\pi r_c$. Next, we introduce a new variable $z = 1 - r_c^2 / r^2$ and use the pole-skipping Matsubara frequencies $\omega=\omega_n = - 2 \pi T i n$ to rewrite the master equation as

\begin{equation}
    4\left(z-1\right)z^2 \psi''(z)+2z \left(3z-2\right)\psi'(z) +\left[n^2-\ell\left(\ell+1\right)z \right]\psi(z)=0. \label{empty_dS}
\end{equation}

Near the cosmological horizon (at $z=0$) we have an ingoing and an outgoing solution,
\begin{equation}
    \psi(z\rightarrow 0) = C_\text{in}z^{-n/2}+C_\text{out}z^{n/2},
\end{equation}
while for $z\to -\infty$ (the location of a stationary observer at the `centre of the universe'),
\begin{equation}
    \psi(z\rightarrow -\infty)=A z^{\ell/2}+B z^{-(\ell+1)/2}.
\end{equation}
In the empty de Sitter space, the equation of motion \eqref{empty_dS} can easily be solved analytically. Setting $C_\text{in}=0$ and expanding around $z\rightarrow -\infty$, we find that the ratio of $B/A$ behaves as
\begin{equation}
    \frac{B(n,\ell)}{A(n,\ell)}\propto\frac{\Gamma\qty(-1/2-\ell)}{\Gamma\qty(1/2+\ell)}\frac{\Gamma\qty(1+\ell+n)}{\Gamma\qty(n-\ell)}. \label{dS_PS}
\end{equation}
For integer $\ell\geq 0$ and integer $n\leq -1$, we therefore find that an infinite tower of pole-skipping points exists at the cosmological horizon for parameters that satisfy the inequality $n \leq - (\ell+1)$. This result can be directly checked by employing the determinant method discussed in Section~\ref{subsec:PS4_systematic}. As in the examples containing black holes, there also exist additional pole-skipping points that are `hidden' from the master function formalism. In the even channel, these pole-skipping points $(\ell,n)$ are $(0,0)$, $(1,1)$ and $(1,-1)$. Finally, in the odd channel, the two additional pole-skipping points are $(0,-1)$ and $(1,0)$.

\section{Discussion and future directions}
\label{sec:Discussion} 

In this paper, we classified and discussed various details of pole-skipping in the backgrounds of massive 4$d$ black holes with maximally symmetric horizons and in spaces with an arbitrary cosmological constant. Linearised gravitational waves in such spaces can be decomposed into two channels: the even and the odd channel of perturbations (with respect to parity). By considering the symmetries of the background, the evolution equations of the gravitational fluctuations decouple, from which one may conclude that the two channels are independent. Nevertheless, at least in 4$d$, this is not the case. There exists a special `integrable' structure that relates the solutions in the two channels: by knowing a solution in one channel, one is able to generate a solution in the other channel. This relation is most easily encoded in the formalism of Darboux transformations, which we generalised to all black hole geometries considered in this paper, and then employed it to discuss the classification of pole-skipping. In particular, owing to the structure of Darboux transformations, we separated the pole-skipping points into a category of algebraically special pole-skipping points (that could be found analytically) and a set of pole-skipping points that is common to both channels. This particular classification is new, rather `natural', and relates the discussion of pole-skipping to a large body of literature in general relativity that had considered algebraically special modes in the past.  

A number of open questions remains to be tackled in the future. One is the question of whether any signatures of pole-skipping can be found in physically accessible gravitational waves. As has now been appreciated for a while, pole-skipping imposes stringent constraints on the structure of gravitational wave Green's functions in momentum space. However, to fully understand its implications, a thorough analysis of pole-skipping in position space may likely be important to find the answer. For the moment, we can comment on the question of indeterminacy of the response. In Fourier (momentum) space, the non-commuting nature of the $\omega$ and $\mu$ limits towards the pole-skipping point implies that Green's functions, and in consequence absorption, transmission and reflection, are not uniquely determined. However, the transformation to position space involves integrals of which the contours can avoid these points (in other words, integrals, or an integral and a sum, over $\omega$ and $\mu$ commute) and lead to unique predictions for absorption and other gravitational wave observables of interest. 

Other interesting research directions include the analysis of pole-skipping through the lens of Darboux transformations for charged and rotating black holes. It would also be interesting to relate pole-skipping to physical effects in those, more complicated types of black hole. 

Finally, despite the fact that the patterns of pole-skipping points in higher dimensions show no apparent commonality between different channel, a particularly interesting open problem is to nevertheless try to understand whether potential hidden, but in some sense similar, new structures exist in the theory of gravitational perturbations of higher-dimensional black holes.

\section{Acknowledgments}
The authors would like to thank Richard Davison and Pavel Kovtun for valuable discussions. S.G. was supported by the STFC Ernest Rutherford Fellowship ST/T00388X/1. The work is also supported by the research programme P1-0402 and the project N1-0245 of Slovenian Research Agency (ARIS). M.V. was supported by the STFC Studentship ST/X508366/1.
\bibliographystyle{jhep.bst}
\bibliography{sn-bibliography}

\end{document}